\documentstyle[emulateapj,psfig]{article}
\newcommand{\be}{\begin{equation}}
\newcommand{\ba}{\begin{eqnarray}}
\newcommand{\ee}{\end{equation}}
\newcommand{\ea}{\end{eqnarray}}  
\newcommand{\etal}{et al.\ }
\def\gtsima{$\; \buildrel > \over \sim \;$}
\def\ltsima{$\; \buildrel < \over \sim \;$}
\def\gsim{\lower.5ex\hbox{\gtsima}}
\def\lsim{\lower.5ex\hbox{\ltsima}}
\def\simgt{\lower.5ex\hbox{\gtsima}}
\def\simlt{\lower.5ex\hbox{\ltsima}}
\def\simpr{\lower.5ex\hbox{\prosima}}

\def\zcr{Z_{\rm cr}}
\def\ekin{\mbox{\cal E}_{\rm kin}}
\def\ekin{{\cal E}_{\rm kin}}
\def\gg{\gamma\gamma}
\def\fgg{f_{\gamma\gamma}}
\def\Lya{Ly$\alpha$~}
\def\sngg{SN$_{\gamma\gamma}$~}
\def\msun{{M_\odot}}
\def\ie{{\frenchspacing\it i.e. }}
\def\eg{{\frenchspacing\it e.g. }}
\def\E3{{\cal E}_{\rm g}^{III}}

\begin{document}
\title{The Detectability of the First Stars and Their Cluster 
        Enrichment Signatures}

\author{E. Scannapieco$^1$, R. Schneider$^{1,2},$ \& A. Ferrara$^3$}
\affil{\footnotesize $^1$ Osservatorio Astrofisico di Arcetri, 
                     Largo E. Fermi 5, 50125 Firenze, Italy}
\affil{\footnotesize $^2$ ``Enrico Fermi'' Centre, Via Panisperna 89/A, 00184 Roma, Italy}
\affil{\footnotesize $^3$ SISSA/International School for Advanced
  Studies, Via Beirut 4, 34014 Trieste, Italy}

\begin{abstract}

We conduct a comprehensive investigation of the detectability of the
first stars and their enrichment signatures in galaxy clusters.  As
the initial mass function (IMF) of these Population III (PopIII) stars
is unknown, and likely to be biased to high masses, we base our
study on analytical models that parameterize these
uncertainties and allow us to make general statements.  We show that
the mean metallicity of outflows from PopIII objects containing these
stars is well above the critical transition metallicity ($\zcr \sim
10^{-4} Z_\odot$) that marks the formation of normal stars.  Thus the
fraction of PopIII objects formed as a function of redshift is heavily
dependent on the distribution of metals and fairly independent of the
mean metallicity of the universe, or the precise value of $\zcr.$

Using an analytic model of inhomogeneous structure formation, we study
the evolution of PopIII objects as a function of the star formation
efficiency, IMF, and efficiency of outflow generation.  For all
models, PopIII objects tend to be in the $10^{6.5}\msun -10^{7.0}
\msun$ mass range, just large enough to cool within a Hubble time, but
small enough that they are not clustered near areas of previous star
formation.  Although the mean metallicity exceeds $\zcr$ at $z \sim
15$ in all models, the peak of PopIII star formation occurs at $z \sim
10$ and such stars continue to form well into the observable range.
We discuss the observational properties of these objects, some of
which may have already been detected in ongoing surveys of
high-redshift Lyman-$\alpha$ emitters.

Finally, we combine our PopIII distributions with the yield models of
Heger and Woosley (2002) to study their impact on the intracluster
medium (ICM) in galaxy clusters.  We find that PopIII stars can
contribute no more than 20\% of the iron observed in the ICM, but if
they form with characteristic masses $\sim 200-260 \msun$, their
peculiar elemental yields help to reconcile theoretical models with
the observed Fe and Si/Fe abundances. However, these stars tend to
overproduce S/Fe and can account only for the O/Fe ratio in the inner
regions of poor clusters.  Additionally, the associated SN heating
falls far short of the observed level of $\sim 1 {\rm keV}$ per ICM
gas particle.  Thus the properties of the first objects may be best
determined by direct observation.

\end{abstract}


\section{Introduction}

The properties of the first cosmic stars born out of a primordial,
metal-free gas may have been quite different from those of later
generations formed from polluted material.  Studies of the
gravitational collapse of the $10^{6-7} M_\odot$ pre-galactic units
that harbor the first stars suggest that these Population III (PopIII)
objects fragmented into clumps with typical masses of $10^{3-4}
M_\odot$ (Abel \etal 1998; Bromm, Coppi, \& Larson 1999, 2002; Abel,
Bryan \& Norman 2000; Bromm \etal 2001) that did not subdivide further
(Omukai \& Nishi 1998; Nakamura \& Umemura 1999; Ripamonti \etal
2002).  As the strong accretion rates (up to 0.1 $M_\odot
\mbox{yr}^{-1}$) onto the central protostellar cores in these clumps
could not have been arrested by radiation pressure, bipolar outflows,
rotation or by any other known mechanism (Ferrara 2001), they are
likely to have evolved into very massive stars.  Such massive star
formation mode would have continued until the gas metallicity passed a
critical threshold of $\zcr \simlt 10^{-4} Z_\odot$ (Omukai 2000;
Bromm \etal 2001; Schneider \etal 2002), which enabled both the
formation of lower mass clumps and some of the above mechanisms to
stop accretion.

This situation leads to the so-called {\it star formation conundrum}
pointed out in Schneider \etal (2002).  If PopIII stars were indeed
very massive $M_\star \ge 100 M_\odot$, stellar evolution models
predict that most of these objects would have evolved into black
holes, which lock in their nucleosynthetic products.  This in turn
would prevent the metal enrichment of the surrounding gas and force
the top-biased star formation mode to continue indefinitely.  It is
only in a relatively narrow mass window in fact, $\Delta_{\gg} \equiv
140 \msun \le M_\star \le 260 \msun$, that ejection of heavy elements
can occur, powered by so called pair-instability supernovae in which
a pair-production instability ignites explosive nucleosynthesis.  
Thus these supernovae may be essential to increase the
intergalactic medium (IGM) metallicity above $\zcr$ and initiate the
transition from top-biased star formation to the formation of
``normal'' (PopII/PopI) stars with a more typical initial mass
function (IMF).

While this transition must have occurred at high redshift in the bulk
of cosmic star formation sites (as suggested by the predicted diffuse
neutrino flux associated to the black hole collapse of the first
stars, Schneider, Guetta, \& Ferrara 2002, and by the interpretation of
the near infrared background, Salvaterra \& Ferrara 2003) massive
PopIII stars are likely to have left some imprints that are observable
today.  One point that might have important observational consequences
is the fact that cosmic metal enrichment has proceeded very
inhomogeneously (Scannapieco, Ferrara, \& Madau 2002; Thacker,
Scannapieco, \& Davis 2002; Furlanetto \& Loeb 2002), with regions
close to star formation sites rapidly becoming metal-polluted and
overshooting $\zcr$, and others remaining essentially metal-free.
Thus, the two modes of star formation, PopIII and normal, must have
been active at the same time and possibly down to relatively low
redshifts, opening up the possibility of detecting PopIII stars.

This is particularly important as metal-free stars are yet to be
detected.  This could be attributed to a statistical paucity of
metal-free stars or may be due to their absence entirely, hinting at
short-lived, massive first stars.  At the same time, theoretical
studies are making rapid progress in understanding the radiative
properties (Tumlinson, Giroux, \& Shull 2001; Bromm, Kudritzki, \& Loeb
2001) and spectral energy distributions of such stars (Schaerer \&
Pell\'o 2002; Oh, Haiman, \& Rees 2001), and their interactions with
the ambient medium (Ciardi \& Ferrara 2001).  Thus, it might well be
that the best approach to detect the first stellar clusters is to
develop careful observation strategies of high-redshift PopIII host
candidates. This is the first aspect that we investigate in detail in
this paper.

However, direct observations are not the only way to learn about the
first cosmic stars. Abundance patterns measured in different cosmic
objects may already show clear imprints of such early star formation.
A prime example are observations of metal-poor halo stars.  Although
these stars cannot be the massive PopIII stars themselves, they might
be old enough to retain some record of the metals ejected by the first
stellar generation (Cayrel 1996; Weiss \etal 2000).  Indeed,
high-resolution spectroscopy of metal poor-halo stars has uncovered a
number of unusual abundance patterns in the iron peak elements in
stars with [Fe/H]$ \sim -2.5$ (McWilliam \etal 1995; Ryan \etal 1996).
In these stars the mean values of [Cr/Fe] and [Mn/Fe] decrease toward
smaller metallicities, [Co/Fe] increases with metallicity, and r- and
s-process elements are nearly absent: trends that cannot be
interpreted with Type II or Type Ia SNa yields (Oh \etal 2001; Umeda
\& Nomoto 2002).  Very recently, the Hamburg/ESO objective prism
survey (HES) team has reported the discovery of a star HE0107-5240
with a mass of $0.8 \msun$ and an iron abundance of [Fe/H] = $-5.3 \pm
0.2$ (Christlieb \etal 2002).  
While this iron value is uniquely low, the overall metallicity of this
star is within the estimated range for $\zcr$ ($10^{-6} Z_{\odot} <
\zcr \leq 10^{-4} Z_{\odot}$, Schneider \etal 2002, 2003) suggesting
that the identification of more stars with iron abundances less than
[Fe/H]$<-5$ will be extremely important to elucidate the conditions
that enable low-mass star formation.

A second indirect approach is to search for PopIII enrichment in the
intergalactic medium itself. Tremendous progress has been made
recently in the determination of the metal content of the IGM both
 observationally (Ellison \etal 2000; Schaye \etal 2000; Songaila 2001)
and theoretically (Gnedin \& Ostriker 1997; Ferrara, Pettini, \&
Shchekinov 2000; Cen \& Bryan 2001; Scannapieco \& Broadhurst 2001;
Madau, Ferrara, \& Rees 2001; Scannapieco, Ferrara, \& Madau 2002;
Thacker, Scannapieco, \& Davis 2002; Furlanetto \& Loeb
2002). However, abundance patterns in the low-column density
Ly$\alpha$ forest can be obtained only for a few elements and are
complicated by uncertain ionization corrections.

In this work, we examine a third possibility for searching for
signatures of PopIII star formation, by studying the intracluster
medium (ICM). Metal abundance experiments are easier in the ICM than
in the more diffuse IGM due not only to higher gas densities, but to the
lower redshift of the targets, the simpler physics of the
collisionally ionized gas, and the larger statistics.  Such an
approach parallels similar efforts by Renzini (1999) and Loewenstein
(2001), but it is specifically designed to test for PopIII star
formation.

The properties of ICM may also provide information as to the energetics
of \sngg, whose high energy output is suggestive of
well-know entropy-floor problem (Loewenstein 2000; 
Lloyd-Davies, Ponman, \& Cannon 2000;
Tozzi \& Norman 2001; Borgani \etal 2001).  Here a wide range of observations have shown that ICM
is much more diffuse and extended than predicted by simple
virialization models, resulting in an X-ray luminosity-temperature
relation that is widely discrepant with simple predictions (Kaiser
1991).  This discrepancy appears to be due to a baseline entropy of
$55 - 110~h^{-1/3}$~keV~cm$^{2}$, where this uncertainty is due to the
unknown extent to which shocks have been suppressed in low-mass
systems.  This excess appears to be distributed uniformly with radius
outside the central cooling regions and provides an additional
constraint on our models of PopIII formation.

Our aim is hence twofold. We first investigate the spatial
distribution and redshift evolution of galaxies hosting PopIII stars.
This study enables us to predict a number of observationally related
quantities (detection probabilities, number counts, equivalent width
distributions) that we combine into a search strategy to find these
systems. Secondly, we turn to the effects of these metal-free stars on
the metallicity patterns and energetics of galaxy clusters.

Throughout this paper, we restrict our attention
to a cosmological model in which $h=0.65$, $\Omega_M$ = 0.3,
$\Omega_\Lambda$ = 0.7, $\Omega_b = 0.05$, $\sigma_8 = 0.87$, $\Gamma
= 0.18,$ and $n=1$, where $h$ is the Hubble constant in units of 100
km~s$^{-1}$~Mpc$^{-1}$, $\Omega_M$, $\Omega_\Lambda$, and $\Omega_b$
are the total matter, vacuum, and baryonic densities in units of the
critical density, $\sigma_8$ is the mass variance of linear
fluctuations on the $8 h^{-1}{\rm Mpc}$ scale, $\Gamma$ is the CDM
shape parameter, and $n$ is the tilt of the primordial power spectrum.
The choice of these parameters is based mainly on measurements of
Cosmic Microwave Background anisotropies (eg.\ Balbi \etal 2000;
Netterfield \etal 2002; Pryke \etal 2002) and the abundance of galaxy
clusters (eg.\ Viana \& Liddle 1996).

The structure of this work is as follows.  In \S2 we derive metal
yields and kinetic energy input from pair-creation and Type II SNe,
and in \S3 we use these quantities to construct a model of metal
dispersal by SN-driven outflows.  In \S4 we present an analytic model
of inhomogeneous structure formation which we use in \S5 to estimate
the distribution of cosmic objects hosting PopIII or PopII/I stars,
accounting for feedback effects due to outflows pre-enriching
neighboring halos.  In \S6 we apply these results to construct the
Lyman-$\alpha$ flux and redshift distribution of PopIII objects and
discuss strategies for their direct detection. In \S7 we turn our
attention to the intracluster medium of galaxy clusters, and relate
the thermal and enrichment properties of this gas to the properties of
PopIII objects.  Conclusions are given in \S8.

\section{Metal Yields and Explosion Energies}

In order to quantify the properties of PopIII and PopII/I objects, we
consider the metal yields and explosion energies of pair-creation 
({SN$_{\gamma\gamma}$) and Type II SNe (SNII),
respectively.  In the PopII/I case we adopt the results of Woosley \&
Weaver (1995) who compute these quantities for SNII as a function of
progenitor mass in the range $12 \msun < M_\star < 40 \msun$ and
initial metallicities $Z=(0,10^{-4},10^{-2},10^{-1}, 1)~Z_{\odot}$.
The corresponding quantities for \sngg are taken from Heger \& Woosley
(2002) assuming progenitors in the mass range $\Delta_{\gg} \equiv 140
\msun \le M_\star \le 260 \msun$.

\subsection{PopII/I Objects}

The total heavy-element mass and energy produced by a given object
depend on the assumed IMF. We define PopII/I objects as those that
host predominantly PopII/PopI stars, \ie objects with initial
metallicity $Z \ge \zcr= 10^{-4} Z_{\odot}$.  For these, we assume a
canonical Salpeter IMF, $\Phi(M) \propto M^{-(1+x)}$, with $x=1.35$,
and lower (upper) mass limit $M_l=0.1\msun$ ($M_u=100\msun$),
normalized so that,
\be
\int_{M_l}^{M_u} dM M \Phi(M)=1.  
\ee 
The IMF-averaged SNII metal
yield of a given heavy element $i$ (in solar masses) is then 
\be
Y_i^{II} \equiv \frac{\int_{\small M_{SNII}}^{\small M_{BH}} dM \Phi(M) M_i }
{\int_{\small M_{SNII}}^{\small M_{BH}}dM \Phi(M)}
\label{eq:yII}
\ee 
where $M_i$ is the total mass of element $i$ ejected by a progenitor
with mass $M$.  The mass range of SNII progenitors is usually assumed
to be $(8 - 100) \msun$.  However, above $M_{BH}=50 \pm 10 \msun$
stars form black holes without ejecting heavy elements into the
surrounding medium (Tsujimoto \etal 1995), and from $(8-11) \msun$ the
pre-supernova evolution of stars is uncertain, resulting in tabulated
yields only between $12\msun$ and $40 \msun$ (Woosley \& Weaver 1995).
To be consistent with previous investigations (Gibson,
Loewenstein, \& Mushotzky 1997; Tsujimoto \etal 1995), in our
reference model (SNII-C in Table 1) we consider SNII progenitor masses
$M_{SNII} = 10 \msun \le M_\star \le M_{BH}=50 \msun$ range, linearly
extrapolating from the lowest and highest mass grid values. Finally,
SNII yields depend on the initial metallicity of the progenitor
star. Here we assume for simplicity that SNII progenitors form out of
a gas with $Z\ge 10^{-2} Z_{\odot}$.  This choice is motivated by the
typical levels of pre-enrichment from PopIII stars (see Section
\ref{Outflows}). Moreover, the predicted SNII yields with initial
metallicity in the range $10^{-4} Z_{\odot} \leq Z \leq 10^{-2} Z_{\odot}$
are largely independent of the initial metallicity of the
star (Woosley \& Weaver 1995).  Different combinations of progenitor
models relevant for the present analysis are illustrated in Table 1.
The last entry in
each row is the number of SNII progenitors per unit stellar mass
formed,
\be
{\cal N}^{II} \equiv \frac{\int_{\small M_{SNII}}^{\small M_{BH}} dM  \Phi(M) }
{\int_{\small M_{l}}^{\small M_{u}}dM M \Phi(M)}.
\label{eq:rateII}
\ee
Finally, we assume that each SNII releases ${\cal E}_{\rm kin}=1.2
\times 10^{51}$~erg, independent of the progenitor mass (Woosley \&
Weaver 1995).

\begin{table*}
\caption{\footnotesize 
The different progenitor models for SNII. The masses are in $\msun$ and ${\cal N}^{II}$ is in units
 of $\msun^{-1}$.}
\begin{center}
\begin{tabular}{@{}|c|cccc|@{}} \hline
SN Model & $M_{SNII}$ & $M_{BH}$ & $Z/Z_{\odot}$ & ${\cal N}^{II}$ \\ \hline
         &            &          &               &            \\[3pt] 
SNII-A   &   12       &   40     &  $10^{-2}$    & 0.00343    \\
SNII-B   &   12       &   40     &    1          & 0.00343    \\
SNII-C   &   10       &   50     &  $10^{-2}$    & 0.00484    \\
SNII-D   &   10       &   50     &    1          & 0.00484    \\  \hline
\end{tabular}
\end{center}
\end{table*}

\begin{table*}
\caption{\footnotesize \sngg Progenitor models. The first three entries are in $\msun$, 
$\, \, {\cal N}^{\gg}$ is in units $\msun^{-1}$ and ${\cal E}_{\rm kin}$ is in units of $10^{51}$~erg. 
The last entry represents the kinetic energy per unit gas mass, ${\cal E}^{III}_{\rm g} = f^{III}_* f_w {\cal N}^{\gg}{\cal E}^{III}_{\rm kin}$ in units of $10^{51}$~erg $\msun^{-1}$ computed assuming $f^{III}_*=0.1$ and $f_w=0.1$
(see text).}
\begin{center}
\begin{tabular}{@{}|c|ccccccc|@{}}\hline
SN Model  &   $M_{C}$  & $\sigma_C^{\rm min}$ & $\sigma_C^{\rm max}$ & ${\cal N}^{\gg}$ & $\fgg$ & ${\cal E}^{III}_{\rm kin}$ & ${\cal E}^{III}_{\rm g}$ \\ \hline
          &            &         &         &                        &                &             &  \\[3pt]
\sngg-A   &   100      &   10    &    17   &$ <6 \times 10^{-5}$   &$[0.03-8]10^{-3}$& $6-7.5$     &$< 4 \times 10^{-6}$ \\
\sngg-B   &   200      &   20    &    50   &$[5 - 4]10^{-3}$       &$0.99-0.8$       & $38.4-37.5$ &$[2-1.5]10^{-3}$  \\
\sngg-C   &   260      &   26    &    70   &$[2.1-2.2]10^{-3}$     &$0.5-0.4$        & $67.1-46.4$ &$[1.4-1]10^{-3}$  \\
\sngg-D   &   300      &   30    &    83   &$[0.4-1.4]10^{-3}$     &$0.09-0.3$       & $70.5-48.2$ &$[2.6-6.6]10^{-4}$  \\
\sngg-E   &   500      &   50    &    149  &$ < 2.2 \times 10^{-4}$&$<0.05$          & $70.6-47.8$ &$<10^{-4}$   \\
\sngg-F   &   1000     &   100   &    314  &$<3 \times 10^{-5}$ &$<6.3 \times 10^{-3}$&$69.9-43.6$ &$1.3\times10^{-5}$\\ \hline 
\end{tabular}
\end{center}
\end{table*}

\begin{table*}
\caption{\footnotesize The IMF-averaged metal yields (in $\msun$) for
          SNII, SNIa, and \sngg.}

\begin{center}
\begin{tabular}{@{}|c|ccccc|c|@{}}\hline
SN Model &   $Y_{\rm O}$   &  $Y_{\rm Si}$ & $Y_{\rm S}$      
& $Y_{\rm Fe}$  &    $Y_{\rm met}$ \\ \hline
         &           &           &            &           &              \\[3pt]
SNII-A   &   1.43    & 0.133     & 0.064      & 0.136     &     2.03     \\
SNII-B   &   1.56    & 0.165     & 0.078      & 0.115     &     2.23     \\
SNII-C   &   1.11    & 0.1       & 0.047      & 0.126     &     1.6      \\ 
SNII-D   &   1.19    & 0.124     & 0.061      & 0.11      &     1.76     \\ \hline 
SNIa     &   0.148   & 0.158     & 0.086      & 0.744     &     1.23     \\ \hline
\sngg-A  & 48.4-47.5 & 2.07-4.37 & 0.559-13.2 &$[0.29-13.3]10^{-2}$ & 59.3 - 61.3 \\
\sngg-B  & 44.2-43.5 & 20.9-18.9 & 8.77-7.89  &5.81-7.41  & 89 - 86.8    \\
\sngg-C  & 37.4-41.8 & 24.3-20.8 & 11.2-9.01  &22.1-11.6  & 104 - 92.4   \\
\sngg-D  & 35-41.4   & 23.6-21.1 & 11.0-9.19  &24.9-12.5  & 104 - 93.4   \\
\sngg-E  & 32.8-41.3 & 22.5-20.8 & 10.5-9.07  &25.7-12.4  & 100 - 92.9   \\
\sngg-F  & 34.3-42.1 & 23.1-19.8 & 10.8-8.51  &24.8-10.6  & 102 - 90.2   \\ \hline

\end{tabular}
\end{center}
\end{table*}

Table 3 shows the IMF-averaged yields for some elements
relevant to our investigation as well as the total mass of metals
released in different SNII models. In the same Table, we also show the
corresponding yields for Type Ia SNe (SNIa), whose values are
mass-independent and have been taken from Gibson, Loewenstein, \&
Mushotzky (1997).

\subsection{PopIII Objects}
\label{ss:popIII}

The detailed shape of the PopIII IMF is highly uncertain, as these
stars have not been directly observed.  In this investigation we adopt
a model that is based on numerical and semi-analytical studies of
primordial gas cooling and fragmentation (Omukai 2000; Bromm \etal
2001; Nakamura \& Umemura 2001; Schneider \etal 2002; Abel, Bryan, \&
Norman 2002), which show that these processes mainly depend on
molecular hydrogen physics.  In particular, the end products of
fragmentation are determined by the temperature and density at which
molecular hydrogen levels start to be populated according to LTE,
significantly decreasing the cooling rate.  The minimum fragment mass
is then comparable to the Jeans mass at these conditions, which is
$\sim 10^3 \msun$, although minor fragmentation may also occur during
gravitational contraction, due to the occasional enhancement of the
cooling rate by molecular hydrogen three-body formation.

Ultimately, in these conditions the stellar mass is determined by the efficiency of clump
mass accretion onto the central protostellar core. For a gas of
primordial composition, accretion is expected to be very efficient due
to the low opacity and high temperature of the gas.  Given the
uncertainties is these models, it is interesting to explore different
values for the characteristic stellar mass $M_C$ and for $\sigma_C$,
the dispersion around this characteristic mass.  In the following we
assume that PopIII stars are formed according to a mass distribution
given by,
\be
\Phi(M) M dM = \frac{1}{\sqrt{2 \pi} \sigma_C} e^{-(M-M_C)^2/2\sigma_C^2} dM,
\ee
where $100 \msun \le M_C \le 1000 \msun$ and $\sigma^{\rm min}_C \le
\sigma_C \le \sigma^{\rm max}_C$. A similar
shape for the initial PopIII IMF has been also suggested by Nakamura
\& Umemura (2001), with a second Gaussian peak centered around a much
smaller characteristic mass of $1 M_{\odot}$, leading to a bimodal
IMF.

The minimum dispersion is taken to be $\sigma^{\rm min}_C = 0.1 M_C$
and the maximum value is set so that less than 1\% of stars form below
a threshold mass of $M_{BH} \sim 50 \msun$, \ie $\sigma^{\rm max}_C =
(M_C-50 \msun)/3,$ such that they will not leave behind small stars or
metal-free compact objects observable today.  With this choice, only
progenitors with mass in the range $\Delta_{\gg}$ are effective in
metal enrichment. The IMF-averaged yields, $Y_i^{\gg}$, and the number
of progenitors per unit stellar mass formed, ${\cal N}^{\gg}$ can be
obtained from expressions equivalent to eqs.\ (\ref{eq:yII}) and
(\ref{eq:rateII}) with the assumed IMF and metal yields from Heger \&
Woosley (2002).  For each pair of $M_C$ and $\sigma_C$ values, we also
compute the parameter
\be
\fgg \equiv \frac{\int_{\Delta_{\gg}} dM M \Phi(M)}
        {\int_{0}^{\infty} dM M \Phi(M)},
\ee
\ie the fraction of PopIII stars ending in 
\sngg. Similarly, the IMF-averaged specific kinetic energy is
\be
{\cal E}^{III}_{\rm kin}
\equiv \frac{\int_{\Delta_{\gg}} dM \Phi(M) E_{\rm kin}}
{\int_{\Delta_{\gg}}dM \Phi(M)}.
\label{eq:eIII}
\ee   
The various models are described in Table 2 and the corresponding yields are given in Table 3.

\section{Galaxy Outflows and Metal Dispersal}
\label{Outflows}

Having constrained the yields and energetics of individual starbursts,
we now relate them to the properties of the resulting SN driven
outflows.  Here we simply model outflows from both PopIII and PopII/I 
objects as spherical shells expanding into the Hubble flow, as in the
formalism described by Ostriker \& McKee (1988) and Tegmark, Silk, \&
Evrard (1993).  These shells are driven only by the internal hot gas pressure
and decelerated by inertia due to accreting material
and gravitational drag while escaping from the host.
As the properties of PopIII objects are quite uncertain, the inclusion
of more detailed physical processes such as cooling,
stochastic variations in the star formation rate (SFR), and external
pressure are not warranted for our purposes here (eg., Scannapieco \&
Broadhurst 2001; Madau, Ferrara, \& Rees 2001; Scannapieco, Ferrara,
\& Madau 2002).  In this case the relevant evolutionary equations
become
\ba
\ddot{R_s} &=& \frac{3 P_b}{\rho R_s}
         - \frac{3}{R_s}(\dot{R_s} - HR_s)^2
                    - \Omega_M  \frac{H^2 R_s}{2}, \nonumber \\
\dot{E_b} &=&  L(t) - 4 \pi R_s^2 \dot{R_s} P_b,
\ea
where the overdots represent time derivatives, the subscripts {\it s}
and {\it b} indicate shell and bubble quantities respectively, $R_s$
is physical radius of the shell, $E_b$ is the internal energy of
the hot bubble gas, $P_b$ is the pressure of this gas, and $\bar \rho$
is the mean IGM background density.  Finally we assume adiabatic
expansion with index $\gamma=5/3$ such that $P_b=E_b/2\pi R_s^3.$

The evolution of the bubble is completely determined by the mechanical
luminosity evolution assigned, $L(t)$.  Again for simplicity, we take
both PopIII and PopII/I objects to undergo a starburst phase of
$t_{\rm SN} = 5 \times 10^7$ years, but with different energy-input 
prescriptions.  We approximately account for the
gravitational potential of the host galaxy by subtracting the 
value of $G M^2 (\Omega_b/\Omega_M) /r_{\rm vir}$, from the total wind
energy, where $M$ is the total mass (dark + baryonic) of the object and $r_{\rm vir}$ is
its virial radius.  Taking the standard collapse overdensity value of
178 this gives a mechanical luminosity of
\ba
L(t) &=&  160 L_\odot M_b \Theta(t_{\rm SN}-t) 
\left[ f_\star^{II,III}  f_w  \ekin {\cal N}^{II,\gg} \right. \nonumber \\
& & - 5 \left.  \times 10^{-12} (M_b h \Omega_M/\Omega_b  )^{2/3}
  (1+z_c) \right] ,  
\label{meclum}
\ea 
where $f_\star$ is fraction of gas converted into stars, $f_w$ is the
fraction of the SN kinetic energy that is channeled into the galaxy
outflow, $M_b$ is the baryonic mass of the galaxy in units of solar
mass, and $z_c$ is the collapse redshift of the object.  Taking a
typical estimate for PopII objects of $f_\star^{II} f_w \ekin {\cal
N}^{II} = 10^{-4}$ (in units of $10^{51} \, {\rm ergs} \, \msun^{-1}$), we
see that the gravitational contribution becomes important for these
objects when $M_b h \Omega_M/\Omega_b \gtrsim 10^{11} (1+z_c)^{-3/2}
\, \msun.$

The only difference in the wind evolution of PopIII and PopII/I arises
from the product, $f_\star^{II,III} f_w \ekin {\cal N}^{II,III}$,
which we define as the ``energy input per unit gas mass'' ${\cal
E}_{\rm g}^{II,III}$, which has 
units of ${10^{51} {\rm erg} \, M_\odot^{-1}}$.
In the PopII/I case, we take $f_\star^{II}$
to be $0.1$, which gives good agreement with the
observed high redshift star formation rates and abundances of metals
measured in high-redshift quasar absorption line systems (Thacker,
Scannapieco, \& Davis 2002; Scannapieco, Ferrara, \& Madau 2002).
Also in this case we constrain $f_w$ by combining the overall
efficiency of 30\% derived for the $2\times 10^8 M_\odot$ object
simulated by Mori, Ferrara, \& Madau (2002) with the mass scaling
derived in Ferrara, Pettini, \& Shchekinov (2000), which was obtained
by determining the fraction of starburst sites that can produce a
blow-out in a galaxy of a given mass.  This gives $f_w(M) =
0.3\delta_B(M)/\delta_B(M=2\times 10^8 M_\odot)$ where
\be 
\delta_B(M)=
\cases{ 
1.0 & $\tilde N_t \leq 1$ \cr 
1.0 - 0.165 \, {\rm ln} (\tilde N_t^{-1}) &
 $1 \leq \tilde N_t \leq 100$ \cr 
[1.0 - 0.165 \, {\rm ln} (100)] \, 100 \, \tilde N_t^{-1}  & 
$ 100 \leq \tilde N_t$ \cr},
\label{eq:db}
\ee 
and $\tilde N_t \equiv 1.7
\times 10^{-7} (\Omega_b/\Omega_M) M/M_\odot$ is a dimensionless 
parameter that scales according to the overall number of SNe produced 
in a starburst, divided by the star formation efficiency, $f_\star^{II}$.

In the PopIII case on the other hand, there are no direct constraints
on either $f_\star^{III}$ or the wind efficiency.  For these objects
we allow these parameters to be free, varying $\E3$ over a large range
as discussed below.  Finally, when outflows slow down to the point
that they are no longer supersonic, our approximations break down, and
the shell is possibly fragmented by random motions. At this point we
let the bubble expand with the Hubble flow.

Using this model, we can compute the comoving radius at a redshift
$z_r$ of an outflow from a PopIII source of total mass $M$
and formation redshift $z_s$, $r_{III}(M,z_s,z_r) \equiv$
$r[\Omega_b/\Omega_M M \E3 ,z_s,z_r]$, and the equivalent quantity for
PopII/I objects, $r_{II}(M,z_s,z_r) \equiv$ $r[\Omega_b/\Omega_M M
{\cal E}_{\rm g}^{II}(M),z_s,z_r].$ In the upper panels of Fig.\ 
\ref{fig:winds}, we plot $r_{III}$ and $r_{II}$ as functions of $z_r$
for sources of various mass scales, choosing two representative $\E3$
values drawn from Table 2.  In the left panels we consider lower
redshift ($z_s=10$) sources with masses $10^7 M_\odot$ to $10^9 M_\odot$
in a model with $\E3 = 10^{-4.0}$, while in the right panels we
consider more-energetic ($\E3 = 10^{-2.5}$), earlier ($z_s=20$) sources
with masses from $10^6 M_\odot$ to $10^8 M_\odot$.  In the PopII/I
case, we take $f_\star \ekin {\cal N}^{II} = 4.84 \times 10^{-4} $,
consistent with models SNII-C and SNII-D, with a 10\% star formation
efficiency and a supernova energy of $\approx 10^{51}$
ergs.  Here we see that each source ejects material to beyond the
radius of the perturbation from which it was formed, and therefore
this material is easily able to affect the formation of neighboring
objects.

In the lower panels of Fig.\ \ref{fig:winds}, we compute the mean
metallicity within the expanding bubbles, for the same range of
parameters as in the upper curves.  For simplicity, we assume that in
the PopIII case, the mass released in metals is equal to twice the
kinetic energy content, that is $Y_{\rm met}/ \msun \approx 2 \ekin /
10^{51} \, {\rm erg}$, which is a reasonable approximation for the
models in Tables 2 and 3.  Throughout this paper, we
adopt a simple model that assumes that half of the metals are ejected
into the outflow in both the PopII/I and the PopIII cases.
Furthermore, we adopt a single metallicity for this material, ignoring
what is likely to be a strong gradient from the bubble to the shell.
Finally, we do not attempt to reproduce the detailed interaction
between SNe and the inhomogeneous ISM of the galaxy, an issue 
studied in detail in Nakasato and Shigeyama (2000).

For the purposes of Fig.\ \ref{fig:winds}, we take $f_w = 1.0$ in all
PopIII objects, which translates into a minimum metallicity in each
outflow, as this gives the largest values of $\E3$ for a fixed
$\ekin$ value.  In the PopII case, metal masses are computed as per
model SNII-C.  The final metallicity of the bubble scales roughly as
$M^{2/5}$ as the total mass in metals in the bubble is $\propto M$
while its final comoving volume is $\propto M^{3/5}$, due to the $r
\propto E^{1/5}$ Sedov-Taylor scaling.  Thus increasing the object
mass by a factor of a hundred increases the final wind metallicity by
almost an order of magnitude, an effect that is enhanced further in
PopII objects by the assumed inefficiency of large outflows, as
parameterized by eq.\ (\ref{eq:db}).  However, even in the smallest
PopIII objects, with the smallest assumed values of $\E3$ and metal
masses, {\em the mean metallicity of the resulting bubble is well
  above the transition metallicity $\zcr$}.  As the parameters in this
figure were chosen to provide a strong lower limit on the metallicity
of these systems, a clear implication is that all enriched regions of
the IGM lie above the critical value necessary for the formation of
PopII/I stars.

Hence, the switch between primordial and recent star formation was
determined primarily by the {\em spatial distribution} of forming
sources and metals in the universe.  This is an essential point in
understanding the properties of primordial enrichment, which is
independent of the details of our models, as the lowest levels of
$Z_{\rm wind}$ derived from our estimates are several times higher
than the most likely values for $\zcr$.  The fraction of PopIII
objects formed as a function of redshift is intimately tied up with
the evolution of the outflows themselves and fairly independent of the
mean metallicity of the universe, or the precise value of the critical
transition metallicity.  Thus, the evolution of PopIII objects can
only be properly tracked from an approach that captures the spatial
distribution of collapsing halos, as such objects are able to form in
pristine areas over a large range of redshifts. Conversely, primordial
objects are able to enrich selected areas of the IGM to levels well
above $\zcr$ and may have made a significant contribution to the
metals found in the biased areas in which galaxies formed most
vigorously.

\section{Spatial Distribution of PopIII objects}


In order to calculate the spatial distribution of star-forming halos,
we employ an analytical formalism that tracks the correlated formation
of objects.  In this model, described in detail in Scannapieco \&
Barkana (2002), halos are associated with peaks in the smoothed linear
density field, in the same manner as the standard Press-Schechter
(1974) approach.  This approach extends the standard method, however,
using a simple approximation to construct the bivariate mass function
of two perturbations of arbitrary mass and collapse redshift,
initially separated by a fixed comoving distance
(see also Porciani \etal 1998).
From this function we can construct the number density of source
halos of mass $M_s$ that form at a redshift $z_s$ at a comoving
distance $r$ from a recipient halo of mass $M_r$ and formation
redshift $z_r$:      
\be
\frac{dn}{d M_s} (M_s,z_s,r|M_r,z_r) =
\frac{\frac{d^2 n}{dM_s dM_r} (M_s,z_s,M_r,z_r,r)}
{\frac{dn}{dM_r} (M_r,z_r)},
\label{eq:biasnum}
\ee where $\frac{dn}{dM_r} (M_r,z_r)$ is the usual Press-Schechter
mass function and $\frac{d^2 n}{dM_s dM_r} (M_s,z_s,M_r,z_r,r)$ is the
bivariate mass function that gives the product of the differential
number densities at two points separated by an initial comoving
distance $r$, at any two masses and redshifts.  Note that this
expression interpolates smoothly between all standard analytical
limits: reducing, for example, to the standard halo bias expression
described by Mo \& White (1996) in the limit of equal mass halos at
the same redshift, and reproducing the Lacey \& Cole (1993) progenitor
distribution in the limit of different-mass halos at the same position
at different redshifts.  Note also that in adopting this definition we
are effectively working in Lagrangian space, such that $r$ is the {\em
initial} comoving distance between the perturbations.  Fortunately as
enrichment is likely to be more closely dependent on the column depth
of material between the source and the recipient than on their
physical separation, this natural coordinate system is more
appropriate for this problem than the usual Eulerian one.  As a
shorthand we define $\frac{dn_{s,r}}{d M_s} (z_s, r) \equiv
\frac{dn}{d M_s} (M_s,z_s,r|M_r,z_r).$

We now wish to compute $f_{III}(M,z)$, the fraction of objects of mass
$M$ forming from gas with $Z < \zcr$ at a redshift $z$, i.e. the sites of
PopIII star formation.  Having computed the fraction of 
halos that are formed from metal-free gas,
 we will then be able to construct the average number of PopIII
and PopII/I outflows impacting a random point in space at a
redshift $z$ as
\ba
\langle N_{0,III}(z) \rangle = 
\int^\infty_{z} d z_s
\int^\infty_{M_{\rm min}(z_s)} d M_s 
\frac{d^2 n}{d M_s dz_s} 
        \, f_{III}\, V_{III}, \\
\langle N_{0,II/I}(z) \rangle = 
\int^\infty_{z} d z_s
\int^\infty_{M_{\rm min}(z_s)} d M_s 
\frac{d^2 n}{d M_s dz_s}
        \, (1-f_{III}) \,V_{II}, \nonumber
\label{eq:n0}
\ea 
where $V_{III} \equiv \frac{4 \pi}{3} r_{III}^3(M_s,z_s,z)$ and
$V_{II} \equiv \frac{4 \pi}{3} r_{II}^3(M_s,z_s,z)$ are the volumes
contained in spheres of radius $r_{III}$ and $r_{II}$ respectively,
$\frac{d^2 n}{dM_s dz} (M,z)$ is the differential Press-Schechter
number density of objects forming as a function of mass and redshift,
and $M_{\rm min}(z_s)$ is the minimum mass that can cool within a
Hubble time at the specified formation redshift $z$, \ie $t_{\rm
cool}(M_s,z_s) \le t_{H}(z_s)$ (see Ciardi \etal 2000).  Finally, we
include a delay between virialization and star formation of $t_{\rm ff}
= (4 \pi G \Omega_0 180 \rho_c)^{-1/2}$ for all objects, which
accounts for the free-fall time.  This is equal to $2.3 (1+z)^{-3/2} h^{-1}$
Gyr in our assumed cosmology.

In constructing these integrals, we divide objects into two regimes,
in terms of their primary cooling mechanism.  In halos with virial
temperatures above $10^4$ K, atomic cooling is effective, and we adopt
the fixed star formation efficiencies of $f_\star^{II}$ and
$f_\star^{III}$ as described above.  In halos with virial temperatures
$ < 10^4$~K, cooling is dependent on the presence of molecular
hydrogen.  In our previous study of PopII/I outflows (Scannapieco,
Ferrara, \& Madau 2002), we assumed in our fiducial model that star
formation in these objects was completely suppressed by ${\rm H_2}$
photodissociation by UV radiation from the first stars.  As this work
focuses on smaller objects, on the other hand, we allow star formation
in such halos, but with a lower overall star formation efficiency.
Thus for both PopIII and PopII/I objects with $T < 10^4$~K, we adopt a
star formation rate of $0.1 f_\star^{II,III}$.  This is physically
motivated by the lower efficiency of molecular hydrogen cooling, which
reduces the number of cooled baryons available to form stars of any
type (Madau, Ferrara, \& Rees 2001).

With these assumptions, we can modify eq.\ (\ref{eq:n0}) to construct
$\langle N(M_r,z_r) \rangle$,
the mean number of outflows impacting a {\it recipient halo} of mass
$M_r$ and formation redshift $z_r$.  In this case we replace
$\frac{d^2 n}{dM_s dz_s} (M,z)$ with the biased number density of
objects within the volume that is able to impact the recipient halo.
As this number is a function of the distance between the source and
recipient halos, $V_{III}$ and $V_{II}$ must be replaced by radial
integrals.  This leads to:
\ba
\langle N(M_r,z_r) \rangle  &=  4 \pi
\int^\infty_{z_r} d z_s
\int^\infty_{M_{\rm min}(z_s)} d M_s \nonumber \\
 & \left[ f_{III} \int^{r_{III}}_0 dr' r'^2 
\frac{d^2n_{s,r}}{d M_s d z_s}(z_s,r') \right. \nonumber \\ 
 & +
\left. \left(1-f_{III} \right)
\int^{r_{II}}_0 dr' r'^2 \frac{d^2n_{s,r}}{d M_s d z_s}(z_s,r')
\right], 
\label{eq:nm}
\ea
where $\frac{dn^2_{s,r}}{dM_s dz} (z_s,r)$ is now the differential
correlated number density of source objects.

Finally, we relate this quantity to $f_{III}(M,z)$. As our formalism
can only account for the correlations between two halos, we must adopt
an approximation for the fraction of objects that are affected by
multiple winds.  The simplest approach is to 
assume that while the winds only impact a fraction of the overall
cosmological volume, their arrangement {\em within that fraction} is
completely uncorrelated.  In this case $f_{III}$ and 
$\left <N \right>$ share the
same relation as they would in the uncorrelated case, \be f_{III}(M,z)
= 1 - \exp[- \left< N(M,z) \right>].
\label{eq:fIII}
\ee
We will adopt this {\em Ansatz} as our fiducial approach throughout
this paper, referring to it as the ``{\em best-guess model}.''

An alternate approximation relates these quantities by assuming that there is
no overlap between outflowing bubbles.  To construct this relation,
let us define $f_1(M,z)$, $f_2(M,z)$,...  as the fraction of objects
at a particular mass scale that are impacted by $1, 2, ... $ many
outflows.  In this case $\langle N(M,z) \rangle = \sum^\infty_{i=1} i
f_i(M,z)$ while $f_{III}(M,z) = 1- \sum^\infty_{i=1} f_i(M,z)
= 1- \left< N(M,z) \right> + \sum^\infty_{i=2}(i-1)
f_i(M,z)$.  Dropping the multiple-wind term gives the
following relation
\be
f_{III}(M,z) = 
\cases{ 1 - \left< N(M,z) \right> & $ \left< N(M,z) \right> < 1$ \cr
        0 & $ \left<  N(M,z) \right>  \geq 1$ \cr}.
\label{eq:fIIIc}
\ee 
Note that as the contribution from
multiple winds is always positive, dropping this term can only decrease
the fraction of PopIII objects in any given model of metal dispersal
and enrichment.  Thus this relation strictly overestimates the fraction of 
objects impacted by winds, and will be referred to as the 
``{\em conservative model}'' below.

With either of these assumptions, eqs.\ (\ref{eq:nm}) and
(\ref{eq:fIII}) or eqs.\ (\ref{eq:nm}) and (\ref{eq:fIIIc}) form a
closed system, which iteratively defines $f_{III}(M,z)$ and $ \left<
N(M,z) \right>$.  Note that as the metallicity of all bubbles are well
above $\zcr$, these equations are completely independent of the
details of the assumed yields of the underlying stellar populations.

\section{The Redshift Distribution of PopIII objects}

Having developed an analytic formalism for PopIII formation, we now
apply this method to study the evolution of primordial objects as a
function of model parameters.  As this evolution depends only on the
spatial distribution of PopIII ejecta, our models can be simply
classified by the combination of parameters entering $\E3$, the total
energy input into outflows per unit gas mass.  To constrain this
combination, we first consider the range of PopIII IMFs as compiled in
Table 2.

Surveying this table, we find that $\ekin^{III} {\cal N}^{\gamma\gamma}
\lesssim 0.2 $ over all choices of $M_c$ and $\sigma_c$, with
much smaller values being found for the majority of models.  As the
star formation efficiency $f_\star^{III}$ is certainly much less than
1, and is more likely to be on the order of 10\% (Scannapieco,
Ferrara, \& Madau 2002; Barkana 2002) while $f_w$ is likely to be
$\lesssim 30 \%$, we can therefore place a firm upper limit of
$ \E3 \leq 10^{-1.5},$
with the most likely range being $\E3 \lesssim 10^{-2.5}       $
(see Table 2). 

In the left panels of Figure \ref{fig:f} we plot the fraction of
PopIII objects, $f_{III}(z)$, in the best-guess model\footnote{We recall that
${\cal E}_{\rm g}^{II,III}$ is expressed in units of ${10^{51} {\rm erg} \, M_\odot^{-1}}$.}
with $\E3 = 10^{-1.5}       $ as a function of mass, as well as for weaker
feedback cases in which $\E3 =10^{-2.5}       $, $10^{-3.5}       $,
and $10^{-4.5}       $. In each model we have taken $f^{II}_\star
{\cal N}^{II} = 4.8\times 10^{-4} \msun^{-1}$, consistent with models
SNII-B and SNII-C with a fixed star formation efficiency of $f_\star =
0.1$, and have calculated eqs.\ (\ref{eq:biasnum}) - (\ref{eq:fIII})
for objects at 25 different mass scales, spaced in equal logarithmic
intervals from $1.0 \times 10^6 M_\odot$ to $1.0 \times 10^{11}
M_\odot$.

In all cases, the large masses are the most clustered and thus
feedback affects them more strongly, resulting in a quicker transition
from primordial to pre-enriched star formation.  In the $\E3
= 10^{-1.5}       $ model, in fact, only objects with masses 
$\sim 1 \times 10^7 M_\odot$ are able to form PopIII stars,
while at higher mass scales the strong feedback of this model
suppresses primordial star formation almost immediately.  At smaller
mass scales, on the other hand, the minimum mass cutoff occurs at very
early times, resulting in the sharp drop seen in the $10^7 M_\odot$
curve.  Thus it is only in a narrow range of masses and redshifts
that PopIII objects are able to be formed in the presence of strong
feedback.

Reducing the energy input helps to ease these restrictions, as can be
seen in the $\E3= 10^{-2.5}       $ and $10^{-3.5}       $ models.  In
the latter case there is a significant population of objects with
masses of a few times $10^8 M_\odot$, while the $\sim 4 \times 10^7
M_\odot$ scale objects continue to be formed down to their $10^4$K
limit at $z \approx 5$.  Even in this model, however, the majority of
PopIII objects tend to be in the $10^{6.5}\msun -10^{7.0} \msun$ mass
range, just large enough to cool within a Hubble time, but small
enough that they are not clustered near areas of previous star
formation.

At even smaller feedback values, the mass and redshift ranges of
PopIII formation widen to the point that they conflict with
observations.  Thus in the $\E3 = 10^{-4.5}       $ model, metal
dispersal is so inefficient that the universe is unable to break
out of the star-formation conundrum.  Pair production SNe are too
infrequent to spread material over a large fraction of the IGM, and
PopIII star formation continues indefinitely.

The same trends are seen in the conservative models, plotted in the
right panels of this figure.  While feedback is more severe in this
case, reducing $\E3$ again results in a systematic widening of the
mass and redshift range of PopIII formation.  As in the previous case,
at the lowest $\E3$ value, the cosmological enrichment become
extremely inefficient, and universe become ensnared in the
star-formation conundrum.  As this model is constructed to
systematically underestimate the fraction of primordial objects, this
result can be taken as a conservative lower bound on feedback.  Thus
we can restrict our attention to feedback models with energy inputs
the range $10^{-1.5}        \ge \E3 \ge 10^{-4.5}       ,$ with a most likely range
of $10^{-2.5}        \gtrsim \E3 \gtrsim 10^{-4.0}       .$

In the left panels of Figure \ref{fig:sfr} we plot the overall star
formation rate per comoving Mpc$^3$ in PopIII and PopII/I objects, for
two choices of $f_\star^{III}$.  We compute these values simply as
\ba
SFR_{II}(t)  &=&  0.1 \frac{\Omega_b }{\Omega_M} 
\int^{10^{12} \msun}_{M_{\rm min}(t)} d M M 
\left[1-f_{III}(M,t)\right] \nonumber \\
& & \times \frac{d^2 n}{dt d M} 
\left\{ 0.1 + 0.9 \theta[M-M_{10^4}(t)]\right\}\\
SFR_{III}(t)  &=&  
f_\star^{III} \frac{\Omega_b }{\Omega_M}
\int^{10^{12} \msun}_{M_{\rm min}(t)} d M M f_{III}(M,t) \nonumber \\
& & \times \frac{d^2 n}{dt d M} 
\left\{ 0.1 + 0.9 \theta[M-M_{10^4}(t)]\right\},
\label{eq:sfr}
\ea
where $\theta$ is the Heaviside step function, which is used to
account for the $0.1 f_\star^{III}$ star formation efficiency assumed
objects with masses smaller than $M_{10^4}(t)$, the mass corresponding
to a virial temperature of $10^4 K$ at the time $t$.  The overall
downturn at $z \sim 4$ in these plots is due to fact that we are only
considering burst-mode star formation, combined with our choice of an
upper mass limit of $10^{12} \msun$, which crudely models the lack of
larger starbursts due to the long cooling times in these
objects. Because quiescent-mode star formation in Pop II/I objects
becomes important below this redshift, we restrict the analysis given
in Figs.\ \ref{fig:sfr} to 8 to redshifts above $4$.  Note
that in all cases PopIII star formation in objects with masses
$\gtrsim 10^{10} \msun$ is negligible, and we impose $f_{III}(M,t) =
0$ for all objects with masses greater than $10^{11} \msun$,
as we do not calculate these values directly.

While the peak of PopIII star formation occurs at $z \sim 10$ in all
models, such stars continue to contribute appreciably to the SFR
density at much lower redshifts.  This is true even though the mean
IGM metallicity has moved well past the critical transition
metallicity.  In the right panels of Figure \ref{fig:sfr}, we show the
average values as computed in our models.  In this case, rather than
fix $f_\star^{III}$, we must now estimate an overall number of \sngg per
unit mass in PopIII objects for each model.  To do this, we assume as
in Figure 1, that $Y_{\rm met}/\msun \sim 2 \ekin / 10^{51} {\rm
ergs}$, and that half the metals are ejected into the IGM.  In this
case $Y_{\rm met} N^{\gg} f_\star^{III} = f_w^{-1} \E3$, and we show our results for
models in which $f_w = 0.3$ and $f_w = 0.1$.  Note that the
metallicities are lower in the $f_w^{III} = 0.3$ models than in the
$f_w^{III}=0.1$ models, as increasing $f_w$ while leaving $\E3$ fixed
corresponds to a decrease in the overall number of \sngg in each
object.  In all cases, however, the average IGM metallicity exceeds
$Z_{\rm cr}$ at very early redshifts of $12-20$, consistent with
previous estimates by Schneider \etal (2002) and Mackey, Bromm, \&
Hernquist (2002).

Unlike these previous investigations, however, our correlated
structure formation model indicates that it is not until much lower
redshifts that outflows are able to spread enriched gas to all regions
in which galaxy formation is taking place.  This inefficiency of
ejection means that appreciable levels of PopIII star formation are
likely to continue well into the observable redshift range of $z
\lesssim 6$, contributing to as much as $10 \%$ of the star formation
rate at $z = 6$, for our most favorable choice of parameters.  Thus it
is likely that the direct observation of PopIII objects is well within
the capabilities of current instruments, and in fact such objects may
have already been seen in ongoing surveys of high-redshift objects.

\section{The Direct Detection of PopIII Stars}

\subsection{Properties of PopIII Objects}

Our findings in \S5 have important implications for the development of
efficient strategies for the detection of PopIII stars in primeval
galaxies.  This is because at any given redshift, a fraction of the
observed objects have a metal content that is low enough to allow the
preferential formation of PopIII stars. As metal-free stars are
powerful Ly$\alpha$ line emitters (Tumlinson, Giroux, \& Shull 2001;
Schaerer 2002; Venkatesan \etal 2003), it is natural to use this
indicator as a first step in any search for primordial objects. This
is even more promising as surveys aimed at finding young, high
redshift systems have already discovered a considerable number of such
emitters (Dey \etal 1998; Weyman \etal 1998; Hu \etal 1999, 2002;
Ajiki \etal 2002; Dawson \etal 2002; Frye, Broadhurst, \& Ben\'\i tez,
2002; Lehnert \& Bremer 2002; Rhoads \etal 2002; Fujita \etal 2003;
Kodaira \etal 2003)

In principle, the calculation of the Ly$\alpha$ luminosity, $L_\alpha$, 
from a stellar cluster is very simple, being directly related to the 
corresponding hydrogen ionizing photon rate, $Q(H)$, by
\be
L_\alpha = c_L (1-f_{\rm esc}) Q(H),
\label{lyaline}
\ee where $c_L \equiv 1.04 \times 10^{-11}$~erg and $f_{\rm esc}$ is
the escape fraction of ionizing photons from the galaxy. As this escape
fraction is relatively uncertain, we adopt here an educated guess of
$f_{\rm esc} = 0.2$, which is based on a compilation of both
theoretical and observational results (see Ciardi, Bianchi, \& Ferrara
2002 for a detailed discussion and references therein). Finally,
$Q(H)$ is computed from evolutionary stellar models.  In this
case, we assume that in all objects, the stars giving rise to the
observed \Lya emission form in a burst, which is an increasingly
reasonable assumption for higher redshift/lower mass systems.  We take
this burst to be coeval with the galaxy formation redshift for all
objects.  For PopIII stars,
we rely on the recent work of Schaerer (2002) who computed the time
evolution of $Q(H)$ in such a burst for several assumed IMFs.  We
consider two models, assuming that that these stars are distributed
according to a Salpeter IMF with a low-mass cutoff equal to either
$M_l = 1 M_\odot$ (``normal'') or $M_l = 50 M_\odot$ (``top-heavy''),
and an upper mass cutoff held fixed at $M_u=500 M_\odot$.  For PopII/I
stars, on the other hand, we make use of the stellar evolution code
STARBURST99 (Leitherer \etal 1999), assuming in all cases that $M_l =
1 M_\odot$ and $M_u=120 M_\odot$ with a metallicity equal to 1/20
$Z_\odot$.

Fig.\ \ref{fig:QH} shows the evolution of the ionizing photon rate as
a function of the time elapsed since the burst in these models.  As
expected, PopIII stars are characterized by shorter lifetimes and much
higher luminosities, principally due to their higher surface
temperatures. Also note the sudden drop of $Q(H)$ at about 40
Myr, corresponding to the lifetime of stars more massive than $\approx 8 M_\odot$. 
For comparison we have also plotted $Q(H)$ for a
$\delta(M_\star=260~M_\odot)$-function IMF, which provides an extreme
example of a bright, short $Q(H)$.  These curves were directly applied
to calculate the \Lya emissivity of PopIII and PopII/I objects as a
function of their stellar mass and redshift according to eq.\
(\ref{lyaline}) and the formalism introduced in \S 5.  It is important
to note that in adopting this approach, we are implicitly ignoring the
destruction of \Lya photons both by dust and by the random motions
of the medium through which they propagate (Neufeld 1991; Haiman \&
Spaans 1999).  To calculate these effects, one must consider a number
of physical properties that are difficult to model, including
metallicity, dust content, gas random motion and column-density
distribution.  Although this has been attempted by Haiman \& Spaans
(1999), we prefer here to avoid these intricacies and consider our
values as upper limits to the actual \Lya emission.

Having calculated the \Lya emission, we can then derive the
probability that a given high-redshift \Lya detection is due to a
cluster of PopIII stars.  Here we allow for the time evolution of the
\Lya line luminosity from the burst down to the redshift at which the
object is observed according to the curves in Fig. \ref{fig:QH}.
Moreover, we assumed, as in eqs.\ (\ref{eq:sfr}) that $f_\star^{II} =
f_\star^{III} = 0.1$ for all objects with masses greater than
$M_{10^4}(t)$ and $f_\star^{II} = f_\star^{III} = 0.01$ for masses
between $M_{\rm min}(t)$ and $M_{10^4}(t).$ Finally, we assume a
top-heavy IMF for PopIII stars, which greatly shortens the average
lifetimes of PopIII systems and boosts their luminosities by almost an
order of magnitude.  The resulting probabilities are displayed in Fig.
\ref{fig:short}.  In this figure, the isocontours in the \Lya
luminosity-redshift plane indicate the probability to find PopIII
objects (\ie galaxies hosting PopIII stars) in a given sample of \Lya
emitters for various feedback efficiencies, parameterized by the value
of $\E3$.  For reference, we also include the available data points
(as described in the caption) and a line corresponding to a typical
detection flux threshold of $1.5 \times 10^{-17}$ ergs cm$^{-2}$
s$^{-1}$.  For the Fujita \etal (2003) data, we have included a random
scatter of $\Delta z = \pm 0.22$ to simulate observational
uncertainties.

In this figure, we see that PopIII objects populate a well-defined
region of the $L_\alpha$-redshift plane, whose extent is governed by
the feedback strength.  Note that the lower boundary is practically
unaffected by changes in $\E3$, as most PopIII objects are in a
limited mass range such that they are large enough to cool
efficiently, but small enough that they are not clustered near areas
of previous star formation.  At lower $\E3$ values, the non-zero
probability area widens considerably: in this case, a smaller volume
of the universe is polluted and PopIII star formation continues at
lower redshifts and in the higher mass, more luminous objects that
form later in hierarchical models.  Above the typical flux threshold,
\Lya emitters are potentially detectable at all redshifts beyond $5$.
Furthermore, the fraction of PopIII objects increases with redshift,
independent of the assumed feedback strength.  For the fiducial case
$\E3 = 10^{-3}$, for example, the fraction is only a few percent at
$z=4$ but increases to approximately $15\% $ by $z=6.$ We then
conclude that the \Lya emission from already observed high-$z$ sources
can indeed be due to PopIII objects, if such stars were biased to high
masses.  Hence collecting large data samples to increase the
statistical leverage may be crucial for detecting the elusive first
stars.

The situation is markedly different, however, if the first stars
formed according to a normal IMF, as shown Fig.\ \ref{fig:long}. In
this case, the chances to detect PopIII stars at low redshifts rapidly
decreases as their flux drops below the sensitivity threshold. This is
particularly evident when the feedback is strong and cosmic metal
enrichment proceeds efficiently.  In this case, detections are far
more difficult, although an observed absence of galaxies hosting
PopIII stars could be used to set lower-limits on the efficiency of
feedback processes in the early universe.

Our simple burst model of high-redshift sources can also be compared
to the observed number counts of \Lya emitters per unit redshift and
square degree. In Fig. \ref{fig:flux}, we plot these number counts for
both for PopII/I objects, which are largely independent of $\E3$, and
PopIII objects whose numbers depend sensitively on this value.  These
curves are computed in both the top-heavy and normal IMF cases and are
contrasted with the data by Hu \etal (1998).
In all cases PopII/I objects dominate the number counts at redshifts
$\lesssim 6$, However, at higher redshift the PopIII objects outnumber
``normal'' galaxies for a heavy IMF, provided the feedback is mild.
In all cases the detection probability increases rapidly with redshift,
as we saw above.

Finally, we show in Fig.\ 8 the cumulative probability distribution
function (PDF) of observed \Lya equivalent widths (EW) characterizing
PopIII objects that fall above the assumed detection threshold of $1.5
\times 10^{-17}$ ergs cm$^{-2}$ s$^{-1}$.  Here we consider a
top-heavy IMF and two models assuming a $1-500 \msun$ Salpeter IMF
with $\log(\E3) = -3.5$ and $\log(\E3) = -4.0$.  Note that the
top-heavy IMF case is independent of both redshift and $\E3$ because
the Ly$\alpha$ luminosity is roughly constant over the short lifetime
of such massive stars, and thus the objects remain detectable over the
full evolution of equivalent widths.  For all models, the equivalent
width PDF does not depend dramatically on the assumed IMF or the
redshift, and PopIII stars are able to populate the large EW ($\log
EW_{Ly_\alpha}> 3$) region of the graph, although such objects are
observed (eg.\ Malhotra \& Rhoads 2002).  This region is impossible to
fill with PopII/I stars, as these stars tend to inflate the low EWs
tail of the PDF. Hence, large EWs are a key indicator of the presence
of primordial stars.

\subsection{Strategies for Detection and  Identification}

The properties identified above can be combined to yield a strategy to
detect the cosmic objects hosting the first stars. Obviously, this
would be a spectacular discovery, as local searches for truly
metal-free stars have been painfully unsuccessful (for a review see
Beers 2000). We propose here to search for these stars in high-$z$
\Lya emitters, where chances might be considerably higher.

The individuated strategy would be as follows: 
{\itemize
\item Collect a large sample of \Lya emitters,
preferably weighted to the highest possible redshifts.

\item Among the objects in this sample, make a priority list ranking
  the possible PopIII candidates on the basis of their luminosity and
  redshift according to the probability distributions shown in Figs
  \ref{fig:short} and \ref{fig:long}.  This can be done modulo the
  choice of an IMF for PopIII stars and a value
  of the feedback strength, but it is not overly dependent on this choice.  In general the higher-redshift objects, close to the
  detection limit are favored in most models.

\item Further select the candidates on the basis of their \Lya EW,
sources showing the largest values receiving  the highest ranking.

\item Return to the most probable candidates and perform extremely 
deep, long exposure spectroscopy.

\item}

The goal is then to find a clear-cut feature that uniquely identifies
a given high-redshift stellar cluster as virtually metal-free.  Here
they key indicator may have been pointed out by Tumlinson, Giroux, \&
Shull (2001), who noted that Population III stars produce large He III
regions, which can emit detectable He II recombination emission.
Recombination lines of He II at $\lambda$1640 \AA (n = 3 $\rightarrow$
2), $\lambda$3203 \AA (n = 5$\rightarrow$ 3), and $\lambda$4686 \AA (n
= 4 $\rightarrow$ 3) are particularly attractive for this purpose
because they suffer minimal effects of scattering by gas and
decreasing attenuation by intervening dust.  Fig.~1 of Tumlinson,
Giroux \& Shull (2001) shows the flux of the He II 1640 line as a
function of source redshift.  For star formation rates of 20
$M_\odot$~yr$^{-1}$ and 5 $M_\odot$~yr$^{-1}$, the $\lambda$1640 flux
is detectable out to $z\approx 5$ at the sensitivity level of current
\Lya emission line surveys, although the flux for $\lambda$4686 is 7.1
times lower.  These lines are uniquely produced with this intensity by
massive PopIII stars.

As an example, let us consider the particular case of the highly
($\approx 30$ times) magnified \Lya emitter detected by Ellis \etal
(2001), located at redshift $z=5.576$.  This system has as an
estimated stellar mass of $10^6 M_\odot$ and a current star formation
rate derived from the  $2 \times 10^{-18}$ ergs cm$^{-2}$ s$^{-1}$ 
Ly$\alpha$ line of 0.5 $M_\odot$~yr$^{-1}$,
implying an approximate age of 2 Myr. No stellar continuum is detected
to an upper limit of $3\times 10^{-20}$~erg~ cm$^{-2}$~\AA$^{-1}$.
For this object, the He II 1640 emission is redshifted roughly into
the Z band (1.078 $\mu$m) where the difficulties of ground-based NIR
spectroscopy make detection challenging, but not impossible.  In this
case, the object falls between the OH sky lines, but it is likely to
be very faint.  The relative flux ratio, $\cal F$ of the \Lya to He II
$\lambda$1640 lines is equal to $17.8 f_{evol}^{-1}$, where $0.4 <
f_{evol} < 2.0 $ is a parameter which accounts for the time evolution
of the stellar ionizing continuum radiation.  Thus the He II
$\lambda$1640 line is about 9-45 times fainter than \Lya,
corresponding to 4-20 $\times 10^{-20}$ ergs cm$^{-2}$ s$^{-1}$,
a difficult experiment, but within in the realm of
possibilities for some existing/planned instruments.  As many high
redshift sources are over an order of magnitude brighter, however, the
discovery space for metal-free galaxies is large and deserves
continued observational emphasis.

\section{PopIII Signatures in the Intracluster Medium}

Our PopIII models can also be directly applied to asses the impact of
the first stars on galaxy clusters, the largest virialized structures
in the universe (see Rosati, Borgani, \&  Norman 2002 for a recent
review).  In hierarchical models of structure formation, clusters are
predicted to form from the gravitational collapse of regions of
several Mpc, associated with rare high peaks in the primordial density
field.  In the process of virialization, a hot diffuse gas is formed,
the intracluster medium (ICM), which represents a substantial fraction
of the cluster mass, $\sim 15\%$. The ICM permeates the cluster
gravitational potential well and emits in the X-ray band via thermal
bremsstrahlung.

Rich clusters of galaxies can be considered to be ``closed boxes'',
isolated systems that reflect the overall history of structure
formation with negligible interaction with the surroundings. In
particular, the ICM should maintain clear imprints of the thermal and
chemical evolution of the baryons and is therefore a promising
location to look for signatures of PopIII stars (see also Loewenstein
2001).  In the following sections we briefly review some of the
available observations of the ICM obtained by the old (ASCA,
Beppo-Sax) and current generations of X-ray satellites (Chandra,
Newton-XMM). Then we present the predictions of the model that we have
developed in the previous sections and finally discuss our results and
compare them with previous analyses.

\subsection{ICM Abundances and Heating: Observational Constraints}

The hot ICM has been extensively studied via X-ray imaging and
spectroscopy, and is observed to be enriched to a significant fraction
of solar metallicity (see \eg Renzini 1997; White 2000 and references
therein).  Although stars represent only about 10\% of the total
baryonic mass of clusters of galaxies (Loewenstein 2000), the favored
mechanism for enriching the ICM is supernovae-driven winds from
elliptical member galaxies. Various galactic wind models have been
proposed (see \eg Fusco-Femiano \& Matteucci 2002; Pipino \etal 2002)
but whether these winds are dominated by the ejecta of SNII or SNIa is
still vigorously debated (Tsujimoto \etal 1995; Gibson, Loewenstein, \&
Mushotzky 1997; Finoguenov, David, \& Ponman 2000; Gastaldello \&
Molendi 2002).  SNIa products are iron-rich whereas those of SNII are
rich in $\alpha$-elements, such as Si and O; therefore it is crucial
to have accurate separate measurements of each of these contributions.

In the following we consider only average elemental abundances and
neglect the detailed spatially-resolved abundance measurements that
can be obtained with Chandra and XMM.  Our reasoning is that the
present study is best suited to make average predictions rather than
applications to specific systems. We consider elemental abundances
that have been averaged over ensembles of clusters with similar
properties.  Abundance measurements are always normalized to solar,
with solar photospheric values from Anders \& Grevesse (1989), and
refer only to the outer regions of clusters to avoid contamination by
central gradients.

A compilation of the available data is shown in Fig.~\ref{fig:R1}.
Triangles indicate ASCA observations and are taken from Table~2 of
Fukazawa \etal (1998).  Dots indicate the results of a recent analysis
of ASCA archival data by Baumgartner \etal (2002).  Thanks to the high
spectral resolution and better sensitivity, XMM-Newton has obtained
the first abundance measurements of oxygen. Recently, Tamura \etal
(2002) have reported new results of abundance measurements and have
consistently analyzed all the available data. The sample consists of
groups and poor clusters, with temperatures in the range [1.2 -
4]~keV. The abundance ratios in the outer regions of the observed
clusters are shown in Fig.~\ref{fig:R1} as solid regions.  The results
for Si and S are consistent with other measurements but there is a
significant change in the O/Fe abundance between the center and the
outer region, varying from 0.5 up to 0.8 solar.

We show in Fig.~\ref{fig:R2} the relative abundances averaged over
three temperature bins, [2 - 4]~keV, [4 - 8]~keV, [8 -
12]~keV\footnote{For each temperature bin, we have averaged over data
points obtained by different groups as shown in Fig.~\ref{fig:R1},
and we have associated only the statistical error with this average,
neglecting any systematic uncertainties.}.  Unfortunately,
measurements of the (O/Fe) abundance are restricted only to the first
bin. However, the relative abundances of the other elements show some
trends: going from the poorer to the richer systems, Fe decreases
whereas (Si/Fe) and (S/Fe) increase. This is consistent with previous
findings (Fukazawa \etal 1998; Finoguenov \etal 2000) and might reveal
a larger role of SNIa in the enrichment of groups and poorer systems
compared with that of rich clusters.  Supersolar (Si/Fe) and
(S/Fe) abundance ratios are observed in richer clusters.

Besides elemental abundances, the observed X-ray
properties of clusters provide strong constraints on the thermal
history of the ICM.  While the optical properties of clusters can be
understood in the context of self-similar models (Kaiser 1986), which
assume pure gravitational heating, the slope of the ICM X-ray
luminosity-temperature ($L_x-T$) relation hints a more complicated
history.  If the ICM was heated only gravitationally, then the X-ray
luminosity should be proportional to $T^2$, yet observations indicate
that this luminosity instead goes as $T^{7/2}.$ At lower temperatures
this relation become even steeper, with clusters of widely different
luminosities all having temperatures $\sim 1$ keV (Cavaliere, Menci, \&
Tozzi 1999).

The accepted explanation for this unusual behavior relies on the
presence of large-scale energy input from astrophysical sources
(Kaiser 1991) prior to the gravitational collapse of the cluster.
This energy increases the ICM entropy, placing it on a higher adiabat
and preventing it from reaching a high central density during
collapse, which decreases its X-ray luminosity (\eg Tozzi \& Norman
2001). For a fixed degree of heating per gas particle, this effect is
more prominent for poorer clusters, whose virial temperatures are
comparable to this extra contribution.  As a result, a $T^{7/2}$
relation is established in hot systems and broken for colder systems,
as observed.  Both semi--analytical studies (\eg Cavaliere, Menci, \&
Tozzi 1998; Wu, Fabian, \& Nulsen 2000; Tozzi, Scharf, \& Norman 2001)
and numerical simulations (\eg Brighenti \& Mathews 2001; Bialek,
Evrard, \& Mohr 2001; Borgani \etal 2001) indicate that $\sim 1$ keV
per gas particle of extra energy is required, with a corresponding
entropy floor of $55 - 110~h^{-1/3}$~keV~cm$^{2}$ (see however, Voigt
\& Bryan 2001).

\subsection{Predictions for PopIII stars in clusters}

Using the distribution of PopIII objects calculated in \S 3 we can
directly relate $\E3$ to the levels of enrichment and heating
observed in clusters.  In order to compare these quantities in a
model-independent way, we work with
$F_{III}$, the fraction of cluster gas that cools into PopIII objects.
While this quantity is dependent only on the quantity $\E3$, it can be easily used to
construct the overall fraction of PopIII stars ($f_\star^{III}
F_{III}$), energy input ($ \ekin^{III} N^{\gg} f_\star^{III} F_{III}$), and
metal yields ($Y_i^{\gg} N^{\gg} f_\star^{III} F_{III}$) in any given
model of PopIII star formation.  

Accounting for the $0.1 f_\star^{III}$ efficiency assumed for
objects with $T_{\rm vir} \leq 10^4$K, we calculate this quantity as
\ba
F_{III}  =  & \frac{1}{\Omega_b \rho_c}
\int^{\infty}_{4} d z_s \frac{\Omega_b}{\Omega_M}
\int^{10^{12} \msun}_{M_{\rm min}(z_s)} d M_s M_s f_{III}(M,z_s)
\nonumber\\
& \times \frac{dn_{s,c}}{dz_s d M_s} 
\left\{ 0.1 + 0.9\theta[M_s-M_{10^4}(z_s)]\right\}
\ea
In this case the relevant recipient halo is the cluster, and all
sources are located within the region of space that later collapses
into the cluster itself.  As $\frac{dn_{s,c}}{dz_s d M_s}(z_c,r)$
coincides almost exactly with $\frac{dn_{s,c}}{dz_s d M_s}(z_r,0)$ if
the distance between the two peaks is less than their collapse radii,
we are free to take the number density at zero separation (Scannapieco
\& Barkana 2002), which is equivalent to the usual progenitor
distribution as first described in Lacey \& Cole (1993).
While these integrals break down for the smallest objects and
redshifts, the rising value of $M_{\rm min}$, coupled with the sharp
drop in $f_{III}(M,z)$ at low redshifts assures that the contribution
from these values is minimal.  Nevertheless we impose an overall
minimum source redshift of $4$ in computing this integral, rather than
include contributions down to the final collapse redshift $z_c$.

In Figure \ref{fig:FIII} we plot $F_{III}(\E3,z_c)$ assuming both the
best-guess (solid) and conservative (dashed) models.  In both cases we
consider clusters with collapse redshifts of $2.0$, $1.5$, $1.0$, and
$0.5$.  Note that these values are completely independent of the mass
of the final cluster, which is typically many orders of magnitude
larger than the masses of PopIII objects.  From the point of view of
PopIII objects, the regions of space that form clusters are equivalent
to closed universes in which the overall matter density has been
increased such that collapse occurs at $z_c$.

A clear implication of this plot is that at most $10\%$ of the gas in
clusters can be cooled into PopIII objects.  This is true even in the
most optimistic case in which $\E3$ is taken to be $1.0^{-4.5}
\msun^{-1}$ and the best-guess model, eq.\ (\ref{eq:fIII}), is used to
relate $f_{III}$ and $\left< N \right>$.  In the more likely range of
values $1.0^{-4.0}        \lesssim \E3 \lesssim 1.0^{-2.5}       $,
$F_{III}$ varies from about $1\%$ to $5\%$ of the cluster gas.  When
combined with the detailed PopIII modeling discussed in \S 2.1, these
limits place strong constraints on the PopIII contribution to ICM
heating and enrichment.

\subsection{ICM Heating}

Our $F_{III}$ values can be simply related to the constraints on
cluster preheating without any further assumptions about the
underlying distribution of stars.  In order to place a firm upper limit
on this contribution, we assume that while only a fraction $f_w$ of
the kinetic energy from \sngg goes into powering outflows and
dispersing metals, the rest of the energy also somehow makes its way
into the ICM.  In this case a total energy input of
\be
{\cal E}^{III}_{\rm kin} N^{\gg} f_\star^{III} F_{III} \, {\rm ergs/\msun}
= 310 f_{w}^{-1} \E3 F_{III} \, {\rm kev/particle}
\ee
is injected into this gas.  
Finally we take a conservative value of $f_{w}=0.1$ for the overall
wind efficiency, such that the energy input into the ICM is large
relative to the outflow energies.  In the lower panel of Figure
\ref{fig:FIII}, we plot this estimate for both the best-guess and
conservative models.  Focusing our attention on the upper curves, we
see that while increasing $\E3$ results in a higher overall level of
heating, the energy input per baryon is always $T_{ICM} \lesssim 0.2
\, {\rm keV/particle}$, even for the maximum reasonable $\E3$ value of
$10^{-1.5}       .$ Thus \sngg heating falls short of the required
level of $\sim 1 {\rm keV}$ per gas particle even in the most
optimistic case.

Furthermore, most of this heating occurs at high redshifts, at which
the mean cosmological density is higher and thus the overall entropy
corresponding to a given temperature is smaller.  Taking a rough value
of the redshift for this energy injection of $z = 8$ (see Fig.\
\ref{fig:sfr}) and ignoring the additional gas density due to the fact
that clusters form from overdense regions gives an overall entropy
level of $ T_{\rm ICM} n_e^{-2/3} \sim 150 h^{-1/3} T_{\rm ICM} \, {\rm
kev \, cm^2}$ for our assumed cosmology.  Even in the most optimistic
heating model, this is barely consistent with the lowest allowed
entropy values of $30 h^{-1/3}$ ~keV~cm$^{2}$.  Thus we conclude that
the energy input from PopIII objects is insufficient to preheat
clusters to their observed levels.  It is important to emphasize that
this result is based simply on energetic arguments, and does not
depend on any assumptions about the underlying PopIII star formation
rate, initial mass function, or properties of \sngg.

\subsection{ICM Metal Enrichment}

In order to estimate the observed (Fe/H), (Si/Fe), (S/Fe) and (O/Fe)
abundances from our models we compute the total mass of a
given element $i$ in the ICM as 
\be
M_i = {\cal N}^{II} M_{\star} \left \{Y_i^{II} + 
Y_i^{Ia} \frac{{\cal N}^{Ia}}{{\cal N}^{II}}+
Y_i^{\gg}\frac{{\cal N}^{\gg}}{{\cal N}^{II}} {\cal R} \right\},
\ee
where $Y_i^{II}$, $Y_i^{Ia}$ and $Y_i^{\gg}$ are the IMF-averaged
yields of the element $i$ (in solar masses) for SNII, SNIa and \sngg
as given in Table~3, ${\cal N}^{II}$ and ${\cal N}^{\gg}$ are the
number of SNII and \sngg per stellar mass formed as given in Tables~1
and 2 and the relative number of SNIa and SNII is left as a free
parameter, which is later constrained by observations.   Finally,
$M_{\star}$ is the total mass of PopII/I stars formed assuming a
Salpeter IMF and ${\cal R} \equiv M_{\star}^{III}/M_\star$ is the
total mass of PopIII stars formed per solar mass of PopII/I stars,
assuming a PopIII IMF as described in section \ref{ss:popIII}. The
abundance ratio of a given element $i$ relative to iron can be
expressed as
\be
(i/{\rm Fe})=\frac{Y_i^{II} + Y_i^{Ia} ({\cal N}^{Ia}/{\cal N}^{II})+
Y_i^{\gg}({\cal N}^{\gg}/{\cal N}^{II}) {\cal R}}{Y_{\rm Fe}^{II} 
+ Y_{\rm Fe}^{Ia} ({\cal N}^{Ia}/{\cal N}^{II})+
Y_{\rm Fe}^{\gg}({\cal N}^{\gg}/{\cal N}^{II}) {\cal R}},
\ee
and therefore depends only on the relative rate of SNIa over SNII and
on the parameter ${\cal R}$.  Conversely, the absolute Fe abundance is
computed as follows, 
\be 
({\rm Fe}/H) = \frac{M_{\rm Fe}}{M_{\rm H}}
=\frac{M_{\rm Fe}}{0.76 M_{\rm gas}} = \frac{0.1 M_{\rm Fe}}{0.76
M_{\star,0}} = \frac{0.1 M_{\rm Fe}}{0.76 (1-R) M_{*}} 
\ee 
where we have assumed that {\it (i)} 76\% of the gas mass is in
hydrogen {\it (ii)} the present-day mass in stars and remnants
$M_{\star,0}$ is 10\% of the gas content {\it (iii)} we have corrected
for the fraction $R$ of the stellar mass of PopII/I stars which is
returned to the gas.  Assuming a Salpeter IMF and remnant masses from
Ferreras and Silk (2000), we find that $R=0.276$.  Thus, the absolute
iron abundance is,
\be
({\rm Fe}/H) = \frac{0.1 {\cal N}^{II}}{0.76 (1-R)}\left \{Y_{\rm Fe}^{II} 
+ Y_{\rm Fe}^{Ia} \frac{{\cal N}^{Ia}}{{\cal N}^{II}}+
Y_{\rm Fe}^{\gg} \frac{{\cal N}^{\gg}}{{\cal N}^{II}} {\cal R}\right\}. 
\label{eq:iron}  
\ee 

Our modeling of $F_{III}$ gives us a
self-consistent estimate for the parameter ${\cal R}$. In particular,
the parameter ${\cal R}$ can be related to $F_{III}$ as, \be {\cal R}
= \frac{M_{\star}^{III}}{M_\star} = \frac{f_*^{III}M^{III}_{\rm
gas}}{(1-R)M_{\star,0}}=\frac{f_*^{III}}{0.1 (1-R)} F_{III}.  \ee For
a given PopIII IMF and efficiency $f_*^{III}$, we can estimate
${\cal E}^{III}_g,$ taking a conservative value of $f_w = 0.1$ (see
the last column of Table~2), and find the corresponding values for
$F_{III}$ and ${\cal R}$. The results are presented in the top panels
of Fig.~\ref{fig:R3}, where we show ${\cal R}$ as a function of ${\cal
E}^{III}_g$ for two possible cluster formation redshifts (0.5 and 2)
for two different choices of the PopIII star formation efficiency (0.1
and 0.5). Each set of horizontal lines indicates the predicted ${\cal
E}^{III}_g$ for \sngg-B (dashed lines), \sngg-C (solid lines) and
\sngg-D (dot-dashed lines) assuming $\sigma_C^{\rm min}$ and
$\sigma_C^{\rm max}$.  The intersection points mark the appropriate
${\cal R}$ for each model.  It can be seen from the figure that ${\cal
R}$ is typically between 1\% and 2\%, and never exceeds $5\%$. This
result is very solid as ${\cal R}$ is largely independent of the
assumed cluster formation redshift and of the detailed PopIII
IMF. Given the range of values for ${\cal R}$ in each \sngg model, we
can estimate the corresponding Fe abundance normalized to solar using
eq.\ (\ref{eq:iron}) and assuming that only PopIII stars contribute to
the ICM metal enrichment. The results are shown in the bottom panels
of Fig.~\ref{fig:R3} using the same line coding as the top panels.
The predicted Fe abundance in model \sngg-C is degenerate in
$\sigma_C^{\rm min}$ and $\sigma_C^{\rm max}$ and therefore appears as
a solid thick line instead of as an extended region. This figure shows
that the typical [Fe/H] abundance that can be contributed by PopIII stars
is at best $\sim 0.07$. Since the observed iron
abundance ranges between 0.2 and 0.4 solar, this means that PopIII
stars can contribute no more than 17\%-35\% of the iron observed in
clusters.

The predicted elemental abundance ratios are shown in
Figs.~\ref{fig:R4} and \ref{fig:R5} as a function of the relative
number of SNIa to SNII. The shaded regions in each panel indicate the
range of observed values for (Fe/H), (Si/Fe), (S/Fe) and (O/Fe)
averaged over the same three temperature bins as in Fig.~\ref{fig:R2}.
In Fig.~\ref{fig:R4} we show the predicted abundances if only SNIa and
SNII enrich the ICM, \ie if ${\cal R}=0$. The different lines
correspond to different progenitor models for SNII (see Table~1).  It
is clear that none of these models can simultaneously account for the
observed abundances, as previously recognized by Loewenstein (2001).
This is true even if one assumes different relative rates of SNIa and
SNII depending on the cluster richness. Recently, Tamura \etal (2002)
claimed that XMM observations of poor clusters were consistent with
enrichment by SNIa and SNII with relative rates in the range 0.2-0.7.
We find, however, that when averaged with ASCA observations at
comparable richness, the agreement is no longer valid as it
overpredicts (S/Fe).

Hereafter, we take as our fiducial model SNII-C and consider whether
PopIII stars with their peculiar nucleosynthetic yields can improve
the agreement with the observed data. The results are shown in Fig.\ 
\ref{fig:R5} where we add a contribution from PopIII stars formed
according to model \sngg-B with $\sigma_C^{\rm min}$. We restrict our
analysis to this model as it provides the best fit to the observed
abundances.  In each panel, abundances are computed assuming the
limiting values for ${\cal R}$ consistent with the model (see
Fig.~\ref{fig:R3}) and are compared with the prediction of model
SNII-C (solid line).

Because of the relatively small contribution of PopIII stars to the
total iron abundance (see Fig.~\ref{fig:R3}), ${\cal N}^{Ia}/{\cal
N}^{II}$ values from $0.3 - 1$ are required to match the observed
data, with higher values corresponding to poorer clusters. However,
pre-enrichment by PopIII stars enhances the Si/Fe ratio and the same
relative rate of Type Ia to Type II SNe can simultaneously account for
(Fe/H) and (Si/Fe). Since S and Si trace each other in \sngg ejecta,
the increase in the Si/Fe ratio is reflected in a similar increase in
S/Fe which tends to be systematically overpredicted with respect to
the observed data. Finally, the O/Fe abundance is always
underestimated for the relative Type Ia over Type II SN rates required
to match Fe and Si/Fe data being 0.4 solar for model \sngg-B. This is
roughly consistent with the value observed by XMM in the inner cluster
regions.

To summarize, PopIII stars can contribute no more than 20 \% of the
iron observed in galaxy clusters but if they form with characteristic
masses of about $200-260 \msun$, their peculiar elemental yields help
to reconcile the observed Fe and Si/Fe abundances. However, they tend
to overpredict S/Fe and can account only for the O/Fe ratio 
in the inner regions of poor clusters.  Additionally, their energy
input is insufficient to heat the ICM to the observed levels.

Observations of elemental abundances will increase in the near future,
thanks to the enhanced sensitivity of the new generation of X-ray
satellites. The presently available data appear to be hard to reconcile
with simple enrichment by SNIa and SNII ejecta and are only marginally
improved by the addition of a phase of pre-enrichment by very massive
PopIII stars.  The observed iron abundance requires, for poor
clusters, very high relative rates of Type Ia over Type II SNe, at
least if their progenitors are formed according to a Salpeter IMF.

\section{Conclusions}

Although the formation of metal-free stars marked the dawn of the
modern universe, their properties remain a cosmological mystery.  As
the cooling rates of collapsing clouds are drastically changed by
even trace amounts of metals, fragmentation in primordial regions is
completely different than those in molecular clouds today.  Similarly,
the processes that regulate protostellar accretion rates in PopII/I
stars failed to operate in primordial gas, radically altering
evolution.  Thus theoretical PopIII star-formation studies, while in
many ways more straightforward than studies of later-forming stars,
can not be directly compared with any observational constraints.  And
although reduced cooling and lack of regulation of accretion both
point to massive star formation, the characteristic mass of such stars
is uncertain to over an order of magnitude.

Observations of PopIII stars have remained equally elusive.  Not a
single metal-free star has been seen in our Galaxy, lending weight to
the top-heavy hypothesis while at the same time forcing observers to
adopt alternative methods. These include relating the elemental
abundance patters observed in metal-poor halos stars with the
properties of the (unobserved) stars that enriched them, an approach
that has uncovered unusual abundance ratios that may point towards
\sngg enrichment. A second method is to examine the metal content of
the high-redshift IGM itself, although such studies are limited by the
small number of observable elements and uncertain ionization
corrections.

In this work, we have explored two alternative methods for studying
PopIII stars: (1) by direct detections of high-redshift sources, which
are accessible to ongoing surveys of Ly$\alpha$ emitters; (2) by
studying the impact of PopIII stars on the intracluster medium, which
has been well observed in the X-ray.  As the properties of the first
stars are extremely uncertain, our investigation has relied on
analytical models that allow us to make the most general statements
possible based on a minimum number of assumptions.

Using a simple 1-D outflow model, we have found that for the full
range of parameters considered, the mean metallicity of outflows from
PopIII objects is well above the critical threshold that marks the
formation of normal stars.  This conclusion is based only on three
assumptions: (1) that $Y_{\rm met}/\msun \sim 2 \ekin/ 10^{51}{\rm
ergs}$ for \sngg as computed by Heger and Woolsley (2002); (2) that
the critical transition metallicity $\zcr$ is less than $10^{-3.5} Z_{\odot}$, a
value favored both theoretically (eg.\ Bromm \etal 2001, Schneider
\etal 2002) and observationally by the observed lack of metal-free
stars in the Galaxy; (3) the spherical outflow model of Ostriker and
McKee (1988).  Thus it is fairly certain that PopIII formation
continued well past the time at which the mean IGM metallicity reached
$\zcr$, as large areas of the IGM remained pristine while others were
enriched to many times this value.  In fact, the history of PopIII
formation is likely to have been determined almost exclusively by the
distribution of high-redshift objects and the efficiency of metal
ejection.

To track the distribution of such high-redshift metals, we made use of
the analytical formalism described in Scannapieco \& Barkana (2002)
(for a similar approach see also Porciani \etal 1998).  This formalism
is a generalization of the standard Bond \etal (1991) derivation of
the halo mass function, and it interpolates smoothly between all
standard analytical limits, including the halo bias expression
described by Mo \& White (1996) and the Lacey \& Cole (1993)
progenitor distribution.  By combining this general tool with an
equally general characterization of the underlying IMF of primordial
stars, we were able to study the development of PopIII objects for a
wide range of characteristic masses of these stars, ranging from 100
to 1000 $\msun.$ This range was then further limited by the simple
observation that PopIII stars are not forming at $z = 0.$ This is
because choosing a characteristic mass in which a large majority of
stars are outside of the $140 \msun \le M_\star \le 260 \msun$ mass
range corresponding to the progenitors of \sngg, results in very
inefficient IGM enrichment, which allows PopIII star formation to
continues indefinitely.  Thus lack of PopIII star formation in the
local universe argues for a PopIII IMF with a characteristic mass of
$200 \msun \lesssim M_c \lesssim 500 \msun$.  Even for the models in
this range, however, the peak of PopIII star formation occurs
relatively late ($z \sim 10$), and such PopIII stars continue to
contribute appreciable to the overall SFR well into the observable
range of $z \lesssim 6$.  In all cases PopIII objects tend to be in
the $10^{6.5} \msun - 10^{7.0} \msun$ mass range, just large enough to
cool within a Hubble time, but small enough that they are not
clustered near areas of previous star formation.

Employing a simple model of burst-mode star formation, we have
developed a rough characterization of the directly observable PopIII
objects.  If PopIII stars formed according to a top-heavy IMF, the
short but bright evolution of such stars would have boosted their
luminosities by over an order of magnitude.  Thus, PopIII objects
would make up a large fraction of the detected high-redshift \Lya
emitters.  In fact, for all top-heavy IMF models considered, we found
that PopIII objects are potentially detectable at all redshifts beyond
5, although the precise fraction was sensitive to the assumed feedback
efficiency.

Such detections would be extremely difficult, however, if PopIII stars
formed according to a normal IMF, for although metal-free stars are
slightly brighter than their PopII/I counterparts (eg.\ Tumlinson,
Shull, \& Venkatesan 2002) , this difference is negligible when
compared to IMF effects.  Thus in this case, PopIII objects would be
detectable only if $\E3$ is small, although the absence of such
galaxies can be used to set lower limits on the efficiency of feedback
processes in the early universe.  We note, however, that the large
equivalent widths seen in the observed populations of \Lya emitters is
well-explained by models containing a high fraction of PopIII stars
(Malhotra \& Rhoads 2001).  If the IMF or feedback efficiency allows
for direct detection, our model suggests the following observational
strategy: (1) collecting a large sample of high-redshift emitters; (2)
selecting those at higher redshifts and closest to the favored flux
range of $1.5 \times 10^{-17}$ ergs cm$^{-2}$ s$^{-1}$; (3) selecting
candidates based on high equivalent widths/lack of continuum
detection; (4) performing deep, long-exposure spectroscopy aimed at
the detection of the 1640\AA \, recombination line of HeII.

Finally, our models allow us to asses the impact of PopIII objects on
the ICM.  We find that regardless of the assumed IMF, no more than
10\% of the gas in clusters could have cooled into PopIII objects.
Thus, contrary to previous studies (eg. Lowenstein 2001), we conclude
that PopIII stars can contribute no more than 20\% of the iron
observed in galaxy clusters.  However, if they formed with
characteristic masses of about $200-260 \msun$, their peculiar
elemental yields help to reconcile the observed Fe and Si/Fe
abundances.  Yet, even in this case, they overproduce sulfur and can
account only for the O/Fe ratio in the inner regions of poor clusters.
Furthermore, their energy input is insufficient to heat the ICM to the
observed levels of 1 keV per particle.

Early searches for PopIII stars have a long history of painful
disappointment, driving astronomers to adopt a wide range of
alternative techniques to explore their imprints on later generations.
Yet the same massive star formation that hid these objects from past
searches may be the key to present day detections.  We may finally be
poised to make our first direct measurements of the primordial
generation of cosmic stars.

\acknowledgments 

We are grateful to E.\ Daddi, S.\ Dawson, B.\ K.\ Gibson, S.\ De
Grandi, and S.\ Molendi, as well as the referee, M. Umemura, for
useful information and fruitful suggestions.  ES was supported in part
by an NSF MPS-DRF fellowship; RS was supported by a grant from the
``Enrico Fermi'' Centre (Italy); we acknowledge partial support from
the Research and Training Network `The Physics of the Intergalactic
Medium' established by the European Community under the contract
HPRN-CT2000-00126 RG29185.

}

\begin{figure}
\centerline{\psfig{figure=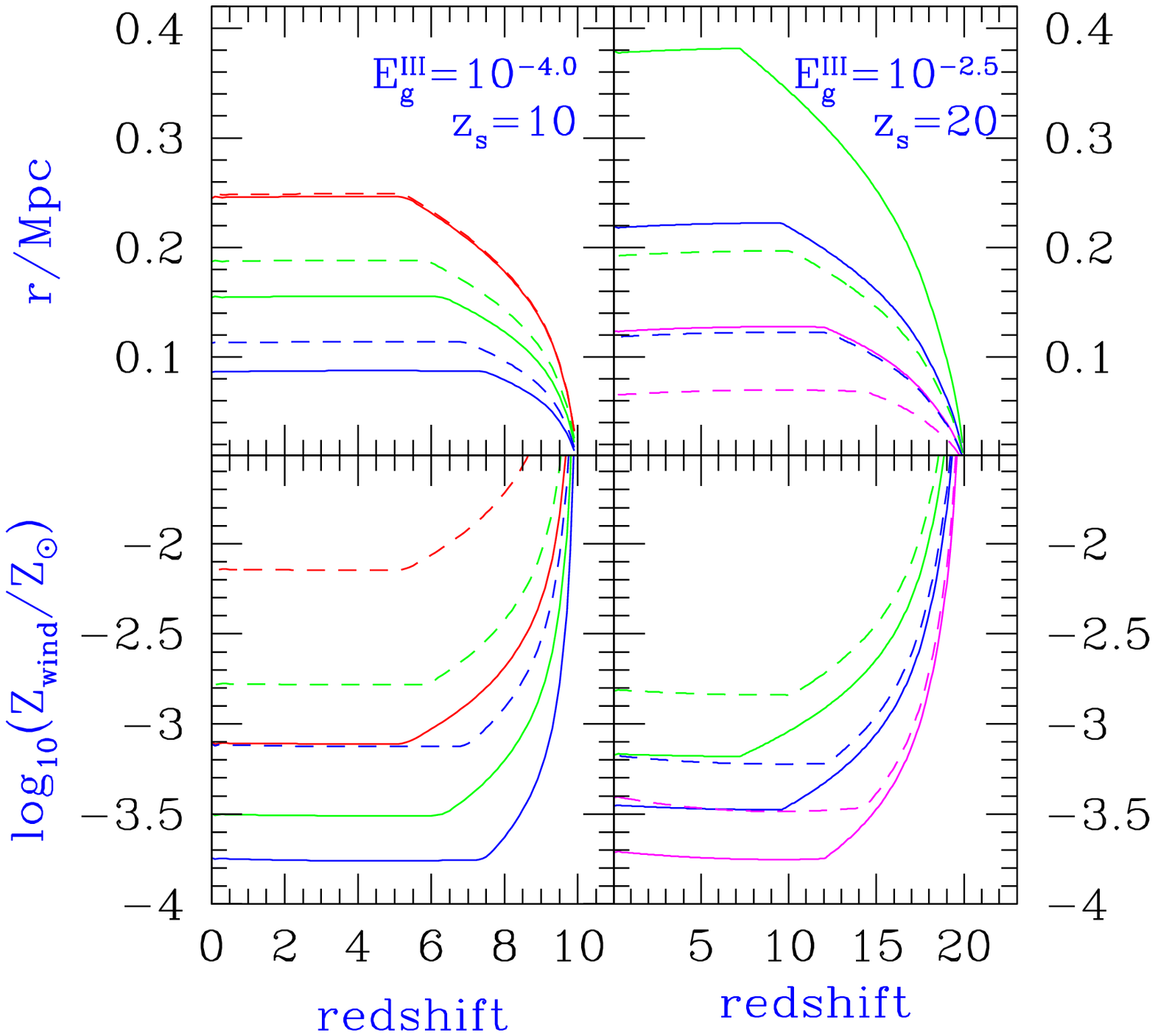,height=18cm}}
\caption{Cosmological metal dispersal.  {\em Top:} Comoving radii
of winds emanating from PopIII (solid lines) and PopII/I (dashed) objects.
In the left panel we take $\E3 = 10^{-4.0}       $ and $z_s = 10$ and consider
sources at three mass scales, taken to be (from top to bottom) $10^9
M_\odot$, $10^8 M_\odot$, $10^7 M_\odot$, corresponding to
fluctuations of initial sizes 190kpc, 90kpc, and 74 kpc, respectively.
In the right panel we take $\E3 = 10^{-2.5}       $ and $z_s = 20$
and consider $10^8 M_\odot$, $10^7 M_\odot$, and
$10^6 M_\odot$ objects, the smallest of which corresponds to
fluctuations with a radius of 19 kpc.  {\em Bottom:} Mean
gas metallicity within each bubble.  The lines represent the same models as
in the upper panels, with $Z_{\rm wind}$ increasing as a function of
source mass.}
\label{fig:winds}
\end{figure}

\begin{figure}
\centerline{\psfig{figure=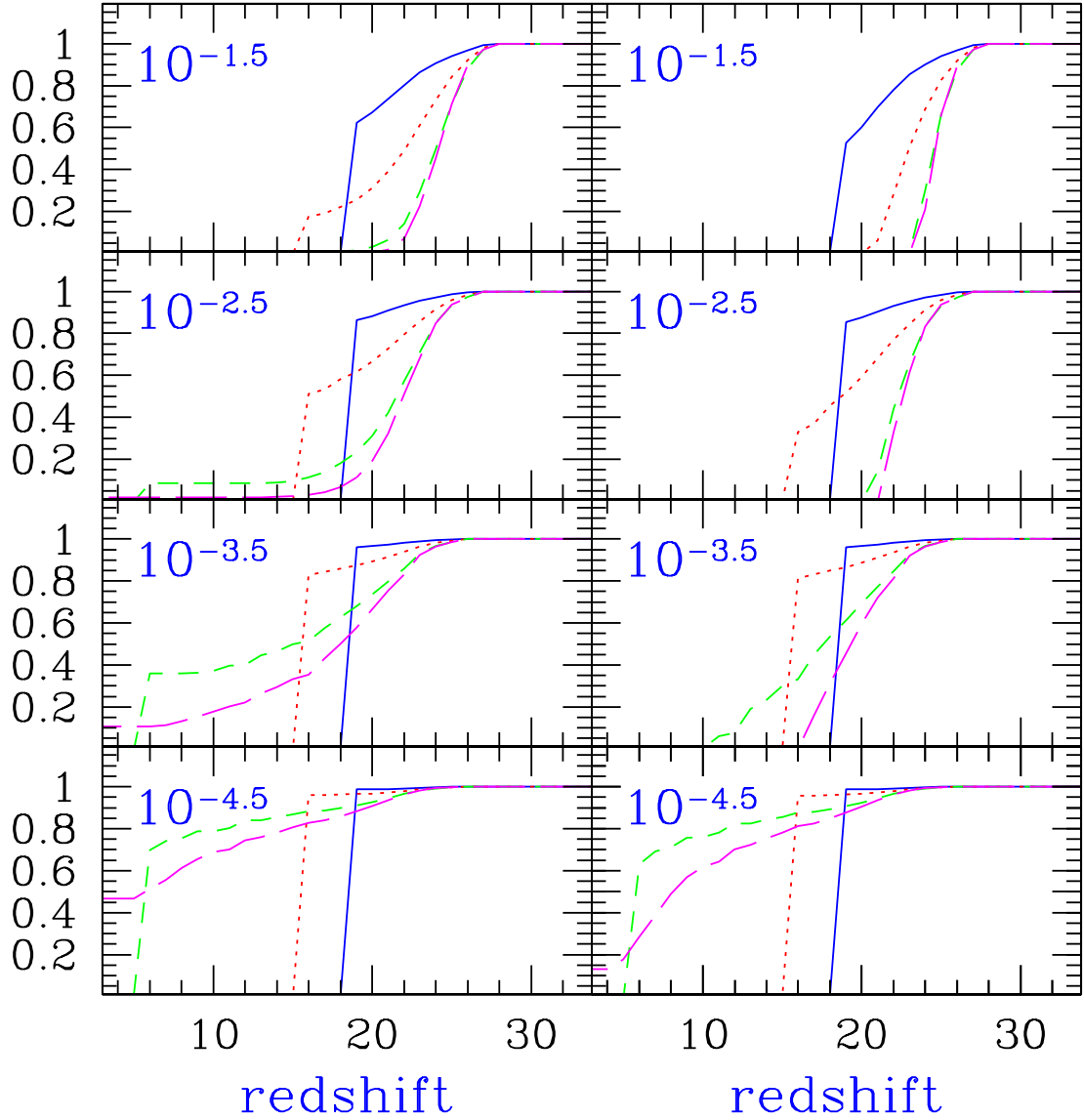,height=18cm}}
\caption{Evolution of PopIII objects.  {\em Left:} Fraction of
PopIII objects for the best-guess models as a function of formation
redshift for various masses and outflow energies.  Each panel is
labeled according to the energy input per gas mass, $\E3.$ In each
panel the solid line, dotted, short dashed, long dashed lines indicate
objects of masses $4.0 \times 10^6 M_\odot$, $1.0 \times 10^7
M_\odot$, $4.0 \times 10^7 M_\odot$, and $1.0 \times 10^8 M_\odot$
respectively.  The sharp drop in $f_{III}$ at the smallest masses is
due to the $M_{\rm min}$ cutoff.  {\em Right:} Fraction of PopIII
objects for the conservative models, curves are the same as in the
left panels.}
\label{fig:f}
\end{figure}

\begin{figure}
\centerline{\psfig{figure=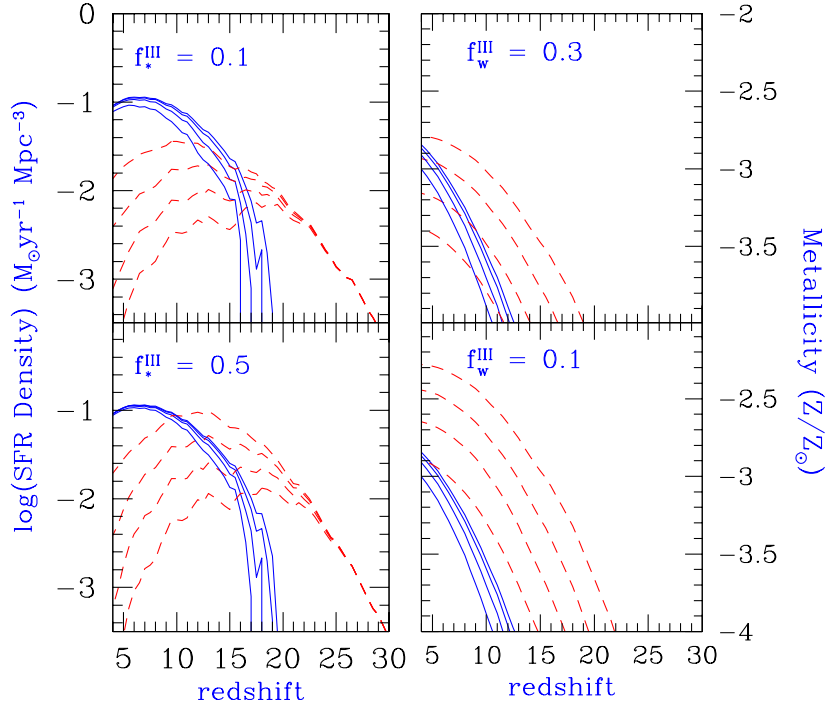,height=9.6cm}}
\caption{Star formation rate densities for PopIII and PopII/I objects.
{\em Top Left :} Star formation rate density per cubic comoving Mpc
for PopIII objects (dashed lines) and PopII/I objects (solid lines).
In this panel we assume $f_\star^{III} = f_\star^{II} = 0.1$ and
consider models in which, from top to bottom, $\log(\E3)$ =
$-4.0 $ $-3.5 $, $-3.0 $, and $-2.5 $.  {\em Bottom Left :} Star
formation rate densities in models in which $f_\star^{III} = 0.5$, for
slightly larger $\log(\E3)$ values of $-3.5 $ though $-2.0.$
{\em Top Right:} IGM average metallicity from PopIII (dashed lines)
and PopII/I (solid lines) objects.  Here $f_w = 0.3$ and 
$\log(\E3)$ ranges from $-4.0 $ to$-2.5$
{\em from bottom to top}, as increasing feedback corresponds
to additional \sngg, which generate more metals. {\em
Bottom Right:} IGM average metallicity with $f_w = 0.1$, lines are as
above.}
\label{fig:sfr}
\end{figure}

\begin{figure}
\centerline{\psfig{figure=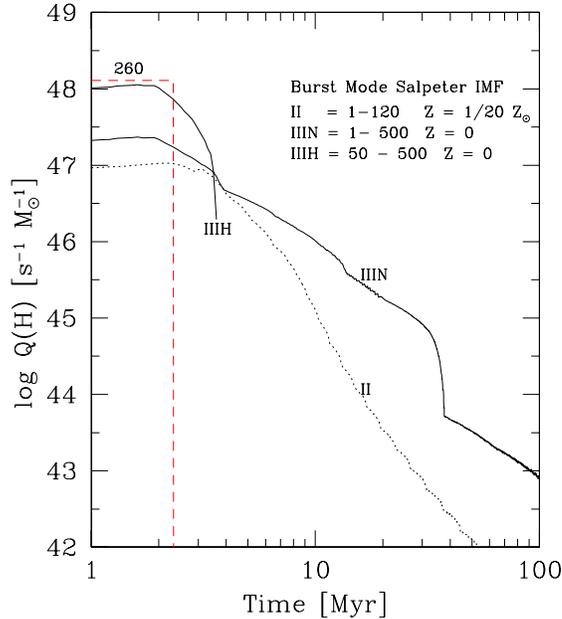,height=8.5cm}}
\caption{Ionizing photon rate per unit stellar mass formed as a function of time in various 
(burst) models of star formation. The dotted line (labeled II) corresponds to PopII/I stars
with metallicity of 0.05 solar and a Salpeter IMF for a range of stellar masses from 
$1-120 \msun$. The solid lines correspond  to PopIII stars.
The extended curve (labeled
IIIN) is for a Salpeter IMF for a range of stars with masses
from $1-500 \msun$, while the peaked curve (IIIH) is for a
``top heavy'' Salpeter IMF, with a range of stars with masses from
$50-500 \msun$.  Finally, the dashed curve is the luminosity per
solar mass for a single metal-free star with a mass of $260 \msun$.}
\label{fig:QH}
\end{figure}

\begin{figure}
\centerline{\psfig{figure=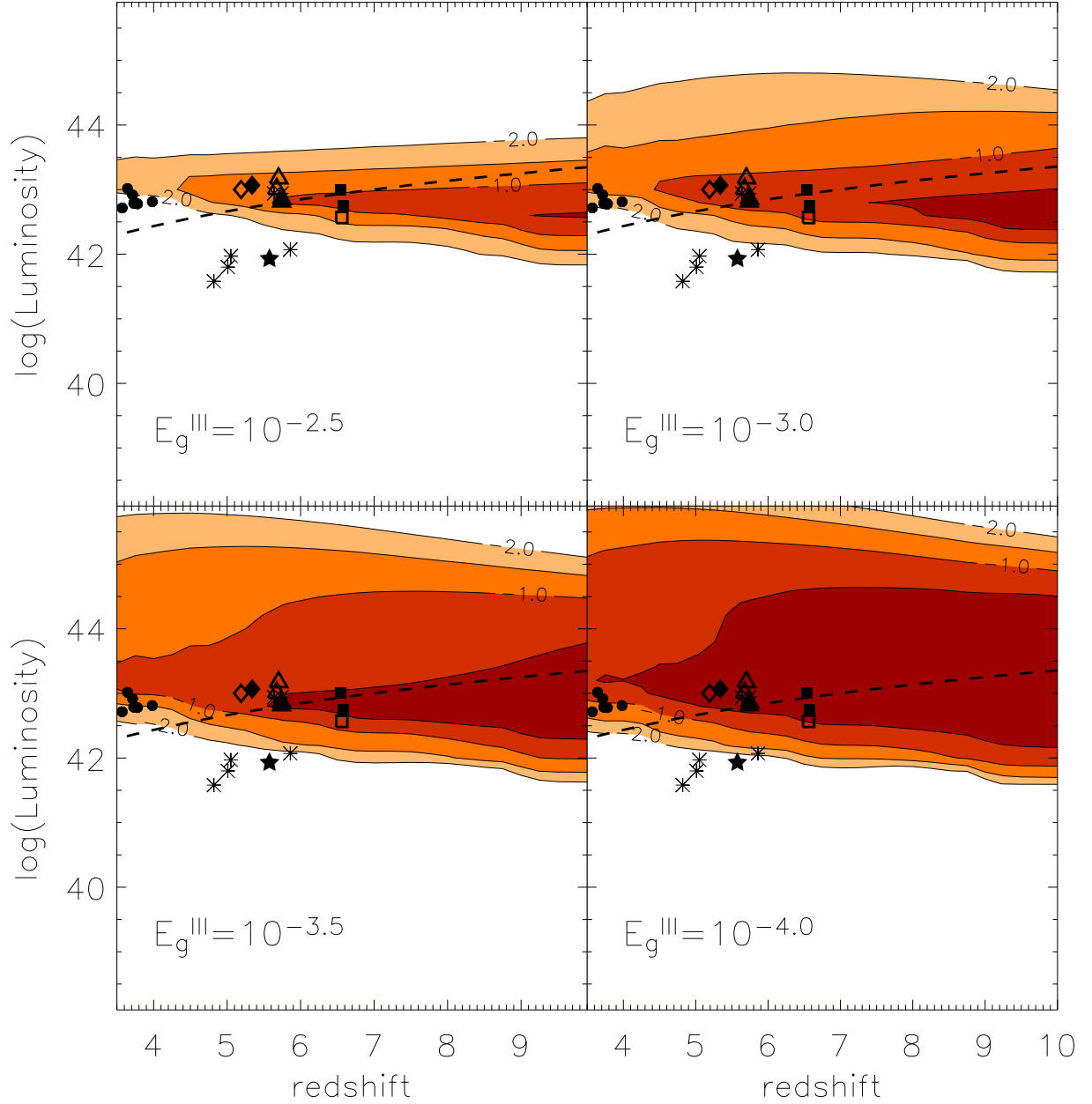,height=18cm}}
\caption{Fraction of PopIII objects as a function of Ly$\alpha$
luminosity and redshift.  Isocontours of fractions $\ge 10^{-2},
10^{-1.5}, 10^{-1}$ and $10^{0.5}$ are shown.
Burst-mode star formation with a
$f_\star^{II} = f_\star^{III} = 0.1$ is assumed for all objects, and
we adopt model $II$ for $Q(H)$ in PopII/I galaxies as given in Figure
\ref{fig:QH}.  In the PopIII case however, we assume a lower cutoff
mass of $50\msun$ as given by model $IIIH$, in this figure.
Each panel is labeled by the assumed $\E3$ value.  
For
reference, the dashed line gives the luminosity corresponding to an
observed flux of $1.5 \times 10^{-17}$ ergs cm$^{-2}$ s$^{-1}$, 
and the various points correspond to observed galaxies.
The filled diamond is from Dey \etal(1998), 
the filled triangle is from Hu \etal (1999), 
the filled star is from Ellis \etal (2001),  
the open diamond is from Dawson \etal (2002), 
the open square is from Hu \etal (2002),  
the asterisks are from Lehnert \& Bremer (2002),
the open triangles are from Rhoads \etal (2002), 
the filled circles are from Fujita \etal (2003), 
and the filled squares are from Kodaira \etal (2003).
Observations by Weyman (1998) and Ajiki (2002) are 
omitted for clarity of presentation.
The curves have been extended slightly past $z=4$ for comparison
with the Fujita \etal (2003) data-set.}
\label{fig:short}
\end{figure}

\begin{figure}
\centerline{\psfig{figure=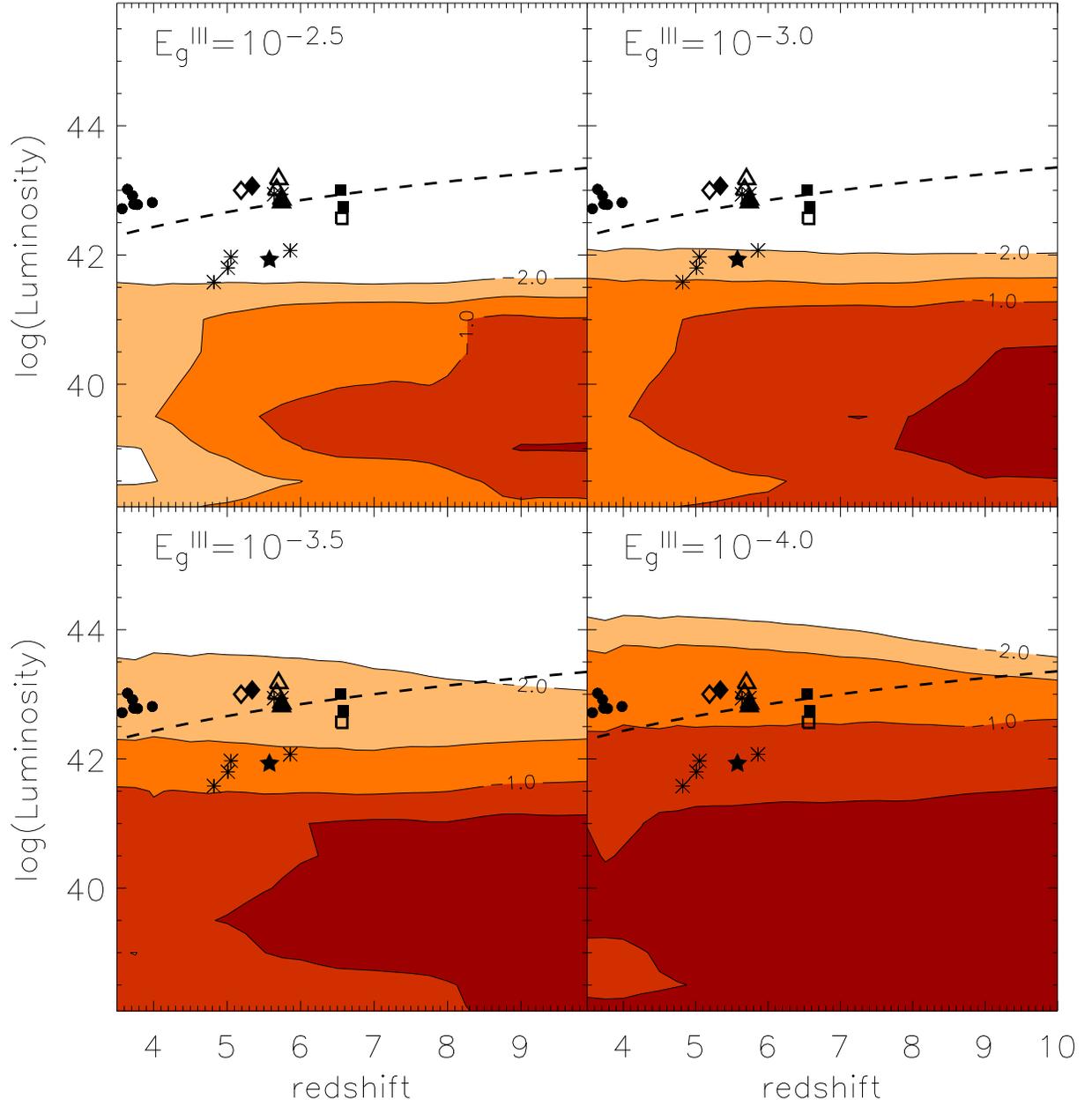,height=18cm}}
\caption{Fraction of PopIII objects as a function of Ly$\alpha$
luminosity and redshift.  In this case a Salpeter IMF with cutoff
masses of $1-500 \msun$ is taken for PopIII objects, greatly
lowering their luminosities.
The dashed lines and points are as in Figure \ref{fig:short}.}
\label{fig:long}
\end{figure}

\begin{figure}
\centerline{\psfig{figure=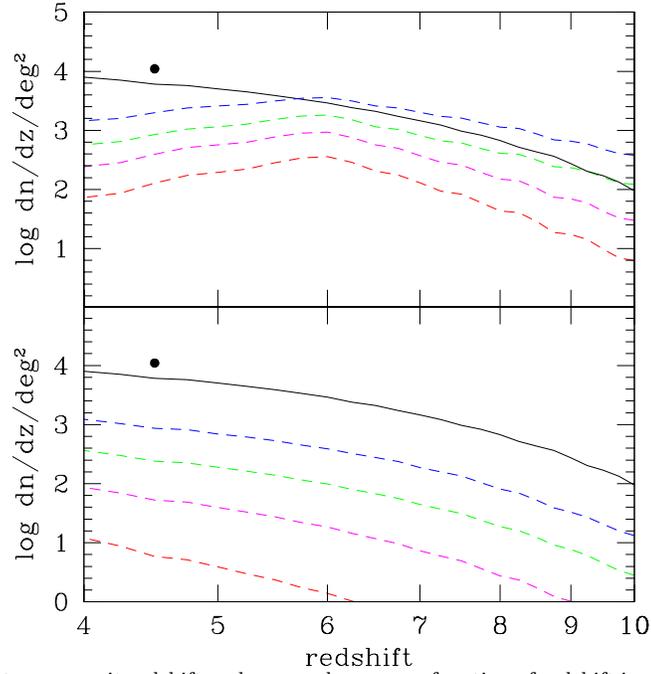,height=9cm}}
\caption{Number counts of Ly$\alpha$ emitters 
per unit redshift and square degree as a function of redshift
in various models of outflow
generation.  {\it Top}: Number counts assuming a top-heavy IMF as in
Fig.\ \ref{fig:short}.  The dashed lines give the total number of Pop
III emitters in models with assumed $\log_{10}(\E3)$ values (from
bottom to top) of $-2.5$, $-3.0$, $-3.5$, and $-4.0$.  The solid line
gives the number counts of PopII/I objects in the $\log_{10}(\E3)=
-3.0$ model, but is largely independent of this value.  A threshold of
$1.5 \times 10^{-17} {\rm ergs} \, {\rm cm}^{-2} {\rm s}^{-1}$ is
assumed in all cases. {\it Bottom}: Curves as in the upper panels, but
assuming a Salpeter IMF as in Fig.\ \ref{fig:long}.
The filled circle gives the value observed by Hu \etal (1998).
}
\label{fig:flux}
\end{figure}

\begin{figure}
\centerline{\psfig{figure=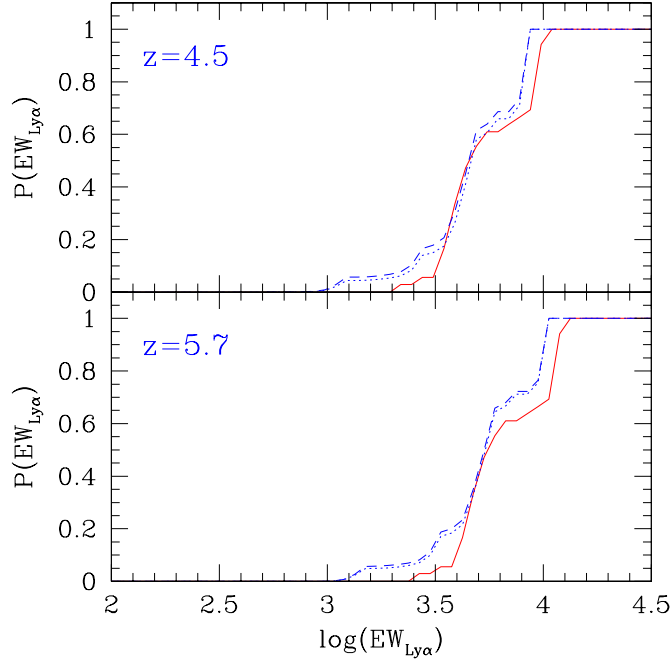,height=9cm}}
\label{fig:ew}
\caption{Distribution of Ly$\alpha$ PopIII observed equivalent widths at two
  characteristic redshifts of $4.5$ (upper panel) and $5.7$ (lower
  panel) for a top-heavy IMF (solid), and two models assuming a $1-500
  \msun$ Salpeter IMF with $\log(\E3) = -3.5$ (dotted) and $\log(\E3)
  = -4.0$ (dashed).  The upper panel in this figure is directly
  comparable to a possible PopIII contribution to the distribution in
  Fig 1.\ of Malhotra \& Rhoads (2002)}
\end{figure}

\begin{figure}
\centerline{\psfig{figure=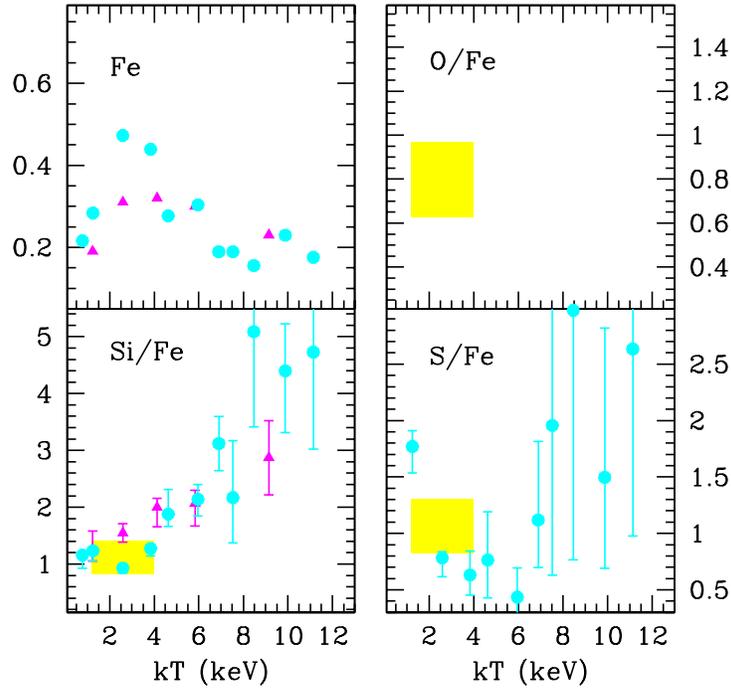,height=10.1cm}}
\caption{Observed Fe abundance and O, Si and S abundance ratios
relative to iron as a function of cluster richness. The data are
normalized to solar photospheric values from Anders \& Grevesse
(1989). Each data point represents an average over clusters with
similar richness. Triangles indicate ASCA observations and are taken
from Table~2 of Fukazawa \etal (1998); dots indicate the results of a
recent analysis of ASCA archival data by Baumgartner \etal (2002); the
solid regions represent the average values observed by XMM (Tamura
\etal 2002).}
\label{fig:R1}
\end{figure}

\begin{figure}
\centerline{\psfig{figure=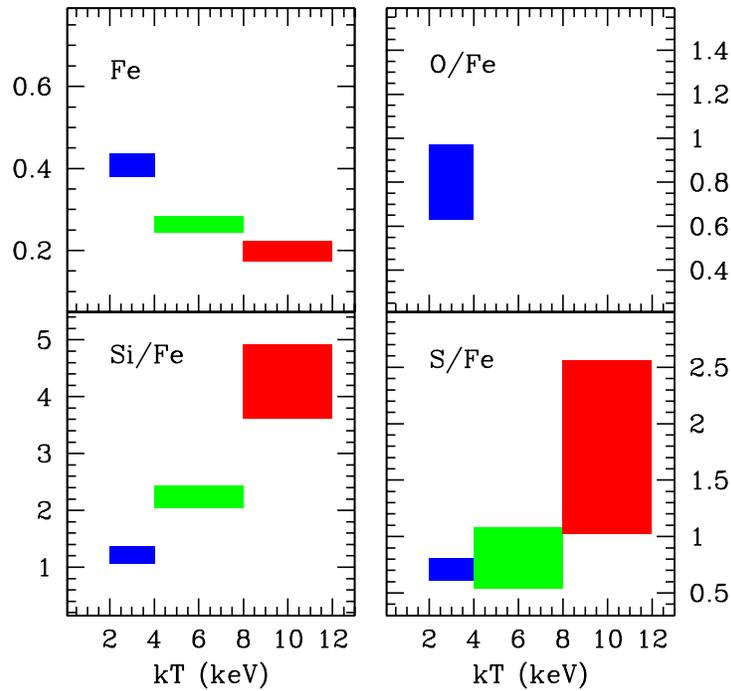,height=10.1cm}}
\caption{Observed Fe abundance and O, Si and S abundance ratios
relative to iron averaged over three cluster richness bins: [2-4]~keV,
[4-8]~keV; [8-12]~keV. The data are normalized to solar photospheric
values from Anders \& Grevesse (1989).}
\label{fig:R2}
\end{figure}

\begin{figure}
\centerline{\psfig{figure=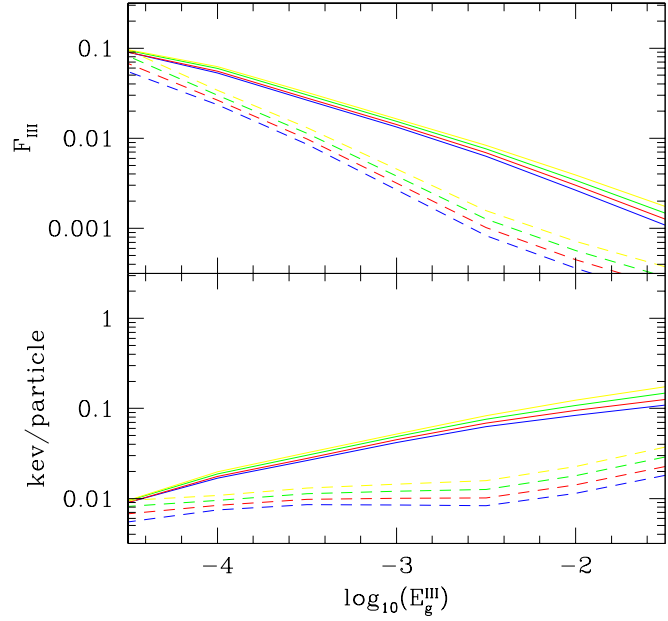,height=8.1cm}}
\caption{{\em Top:} Fraction of cluster gas mass that has collapsed
into PopIII objects as a function of energy input;  solid
lines correspond to the best-estimate model, and the dashed lines
correspond to the conservative model.  In each model the lines
indicate (from top to bottom) clusters with collapse redshifts of 2,
1.5, 1, and 0.5 respectively.  Even in the lowest-energy, best-estimate
model at most 10\% of the cluster gas is processed by PopIII objects.
{\em Bottom:} An optimistic estimate of the level of ICM heating due
to PopIII outflows.  Here we take $f_w = 0.1$ for all values of $\E3$
but assume that all of the supernova energy $\E3 f_w^{-1}$, is
injected into the ICM.  Even in this case the energy input in all
models falls far short of the observed 1 kev/particle level of
preheating.}
\label{fig:FIII}
\end{figure}

\begin{figure}
\centerline{\psfig{figure=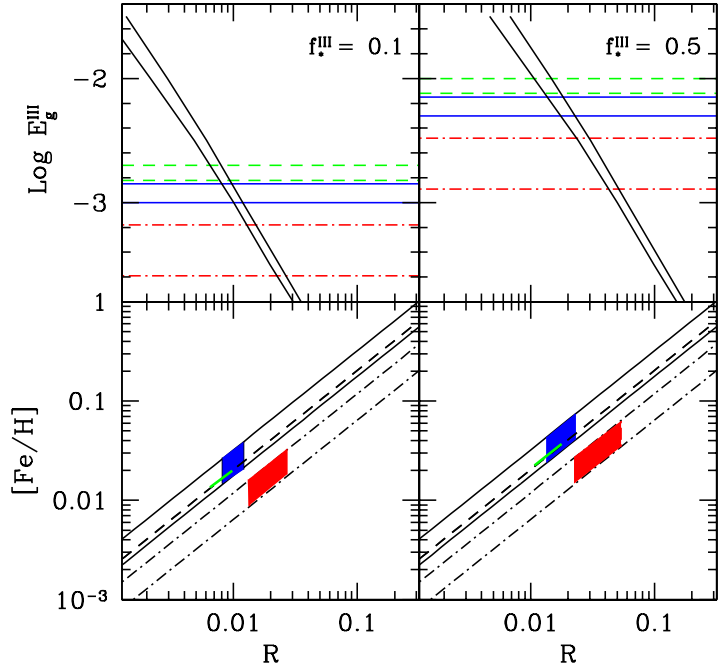,height=8.8cm}}
\caption{{\it Top Panels}: Predicted ratio ${\cal R}=
M^{III}_{\star}/M_{\star}$ as a function of the energy per unit gas
mass ${\cal E}^{III}_{\rm g}$ assuming clusters formation redshifts
0.5 (lower curve) and 2 (higher curve). Horizontal lines indicate the
predicted ${\cal E}^{III}_{\rm g}$ for models \sngg-B (dashed),
\sngg-C (solid) and \sngg-D (dot-dashed) with values for $\sigma_C^{\rm min}$ and
$\sigma_C^{\rm max}$ as given in Table 2 assuming two different values for the PopIII star
formation efficiencies: $f^{III}_*=0.1$ ({\it left panel}) and
$f^{III}_*=0.5$ ({\it right panel}). The intersection with the
predicted ${\cal R}= M^{III}_{\star}/M_{\star}$ indicate the range of
values for this parameter appropriate for each model. {\it Bottom
Panels}: Predicted iron abundance in different \sngg models (with
the same line coding as in top panels) as a function of ${\cal R}$.
The solid regions indicate the range of Fe abundances that might be
achieved in different models assuming the appropriate range of values
for ${\cal R}$ (see text). Left and right panels reflects the two
different star formation efficiencies.}
\label{fig:R3}
\end{figure}

\begin{figure}
\centerline{\psfig{figure=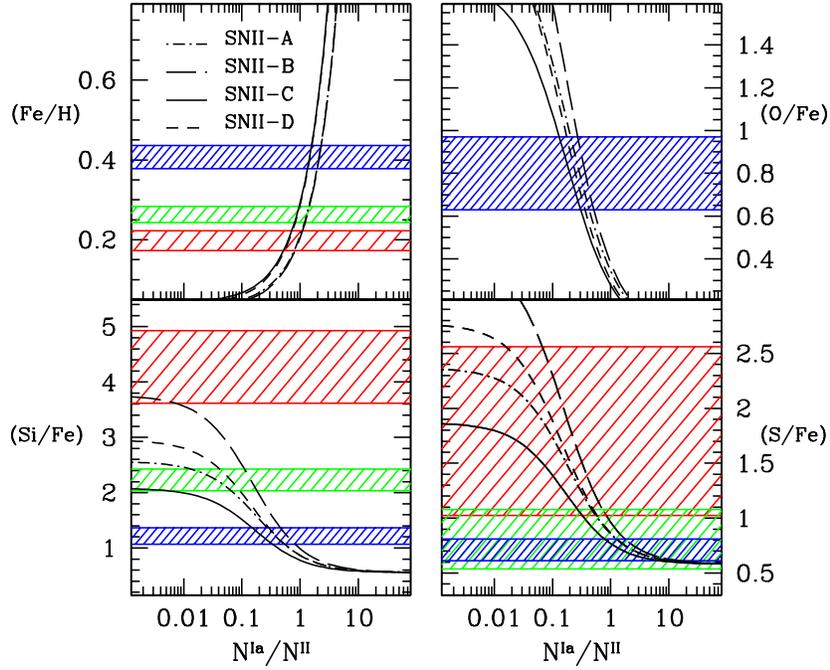,height=9cm}}
\caption{Predicted ICM (Fe/H), (Si/Fe), (O/Fe) and (S/Fe) abundances
(in solar values) as a function of the relative number of SNIa and
SNII. Dashed regions indicate the range of observed values in three
cluster richness bins as in Fig.~\ref{fig:R2} and the different lines
correspond to different SNII progenitor models (see Table~1).}
\label{fig:R4}
\end{figure}

\begin{figure}
\centerline{\psfig{figure=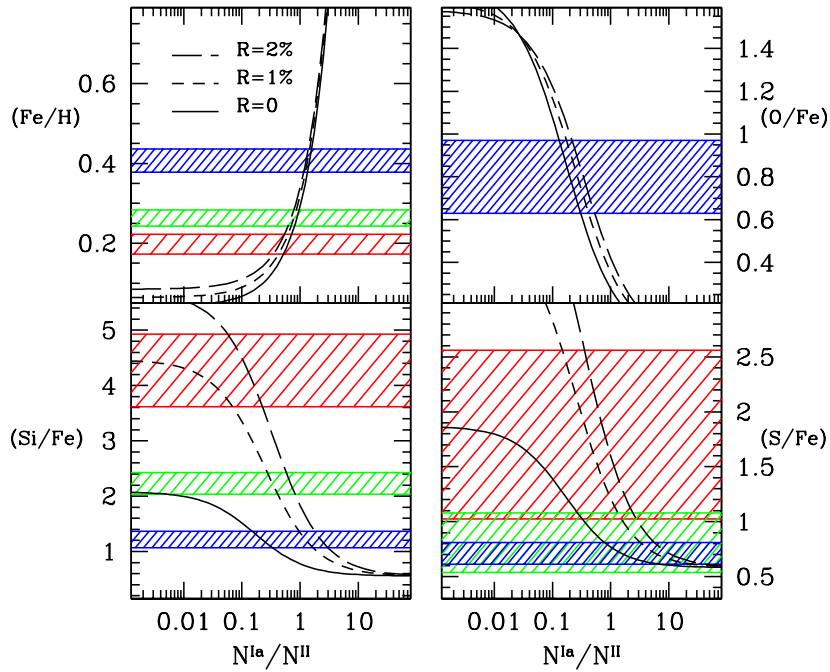,height=9cm}}
\caption{Predicted ICM (Fe/H), (Si/Fe), (O/Fe) and (S/Fe) abundances
(in solar values) as a function of the relative rate of Type Ia over
Type II SNe assuming model \sngg-B for PopIII stars and model SNII-B
for Type II SNe. The long and short-dashed lines correspond to the
limiting values for ${\cal R}$ allowed by the model (see
Fig.~\ref{fig:R3}). For comparison, the solid line shows the predicted
abundances if only Type Ia and Type II SNe contribute to the ICM
enrichment. Dashed regions indicate the range of observed values in
three cluster richness bins as in Fig.~\ref{fig:R2}.}
\label{fig:R5}
\end{figure}

\end{document}